%% file: HIG-18-015_temp.tex
\pdfoutput=1

\documentclass[11pt,twoside,a4paper,cmspaper,final,collab]{cms-tdr}
\def\svnVersion{934769e-D}\def\svnDate{2020/07/28}\def\cmsCernNoTag{CERN-EP-2019-277}\def\cmsCernDate{\today}\def\cmsMessage{"Published in the Journal of High Energy Physics as \href{http://dx.doi.org/10.1007/JHEP07(2020)126}{\doi{10.1007/JHEP07(2020)126}.}"}

\begin{document}\cmsNoteHeader{HIG-18-015}

\hyphenation{had-ron-i-za-tion}
\hyphenation{cal-or-i-me-ter}
\hyphenation{de-vices}

\newlength\cmsFigWidth
\ifthenelse{\boolean{cms@external}}{\setlength\cmsFigWidth{0.49\textwidth}}{\setlength\cmsFigWidth{0.65\textwidth}}
\providecommand{\cmsLeft}{left\xspace}
\providecommand{\cmsRight}{right\xspace}
\providecommand{\cmsTable}[1]{\resizebox{\textwidth}{!}{#1}}

\newcommand{\mc}[1]{\multicolumn{1}{c}{#1}}
\newcommand{\refsec}[1]{Section~\ref{sec:#1}\xspace}
\newcommand{\reffig}[1]{Fig.~\ref{fig:#1}\xspace}
\newcommand{\reftab}[1]{Table~\ref{tab:#1}\xspace}
\newcommand{\refcite}[1]{Ref.~\cite{#1}\xspace}
\newcommand{\refcites}[1]{Refs.~\cite{#1}\xspace}
\newcommand{\massRangeAssCh}{0.2 to 3\TeV}
\newcommand{\massRangeSCh}{0.8 to 3\TeV}
\newcommand{\xsecLimitAssComb}{9.25 to 0.005\unit{pb}}
\newcommand{\xsecLimitAssCh}{21.3 to 0.007\unit{pb}}
\newcommand{\xsecLimitSCh}{4.5 to 0.023\unit{pb}}
\newcommand{\sqrts}[1]{\ensuremath{\sqrt{s}=#1\TeV}\xspace}
\newcommand{\intLumi}{35.9\fbinv}
\newcommand{\pb}{\unit{pb}}
\newcommand{\HTtrig}{\ensuremath{\HT^{\text{trig}}}\xspace}
\newcommand{\pTtrig}{\ensuremath{\pt^{\text{trig}}}\xspace}
\newcommand{\tautau}{\ensuremath{\PGtp\PGtm}\xspace}
\newcommand{\cbbar}{\ensuremath{\PQc\PAQb}\xspace}
\newcommand{\csbar}{\ensuremath{\PQc\PAQs}\xspace}
\newcommand{\bsgamma}{\ensuremath{\PQb\to\PQs\gamma}\xspace}
\newcommand{\TopQuark}{top quark\xspace}
\newcommand{\TopAntitopQuark}{top quark-antiquark\xspace}
\newcommand{\btagging}{\PQb tagging\xspace}
\newcommand{\bjet}{\mbox{\PQb jet}\xspace}
\newcommand{\toptaggingRes}{\ensuremath{\PQt^\text{res}}\xspace}
\newcommand{\btagged}{\PQb-tagged\xspace}
\newcommand{\boostedAnalysis}{boosted analysis\xspace}
\newcommand{\tjet}{\PQt jet\xspace}
\newcommand{\Wjet}{\PW jet\xspace}
\newcommand{\resolvedAnalysis}{resolved analysis\xspace}
\newcommand{\GenuineB}{Genuine \PQb jets\xspace}
\newcommand{\mHpm}{\ensuremath{m_{\PSHpm}}\xspace}
\newcommand{\mSD}{\ensuremath{m_{\text{SD}}}\xspace}
\newcommand{\mh}{\ensuremath{m_{\Ph}}\xspace}
\newcommand{\PAA}{{\HepParticle{A}{}{}}\Xspace}
\newcommand{\mA}{\ensuremath{m_{\PAA}}\xspace}
\newcommand{\mhonetwentyfive}{\ensuremath{M_{\Ph}^{125}(\PSGc)}\xspace}
\newcommand{\tanbeta}{\ensuremath{\tanb}\xspace}
\newcommand{\sPPtoHptb}{\ensuremath{\sigma_{\Pp\Pp\!\to\PSHp\PAQt(\PQb)}}}
\newcommand{\sPPtoHmtb}{\ensuremath{\sigma_{\Pp\Pp\!\to\PSHm\PQt(\PAQb)}}}
\newcommand{\sPPtoHtb}{\ensuremath{\sigma_{\PSHpm\PQt(\PQb)}}}
\newcommand{\sPPtoHpm}{\ensuremath{\sigma(\Pp\Pp\!\to\PSHpm)}}
\newcommand{\Hptb}{\ensuremath{\PSHp\!\to\!\PQt\PAQb}}
\newcommand{\BHptb}{\ensuremath{\mathcal{B}(\PSHp\to\PQt\PAQb)}}
\newcommand{\BHmtb}{\ensuremath{\mathcal{B}(\PSHm\to\PAQt\PQb)}}
\newcommand{\BHtb}{\ensuremath{\mathcal{B}(\PSHpm\to\PQt\PQb)}}
\newcommand{\pseudorapiditySymbol}{\ensuremath{\eta}\xspace}
\newcommand{\absEta}{\ensuremath{\abs{\pseudorapiditySymbol}}\xspace}
\newif\ifMC\MCfalse
\newcommand{\MC}{\ifMC MC\else Monte Carlo (MC)\MCtrue\fi\xspace}
\newif\ifHDM\HDMfalse
\newcommand{\TwoHDM}{\ifHDM 2HDM\else two Higgs doublet model (2HDM)~\cite{Gunion:2002zf,Akeroyd:2016ymd,Branco:2011iw,Craig:2012vn}\HDMtrue\fi\xspace}
\newcommand{\TwoHDMs}{\ifHDM 2HDMs\else two Higgs doublet models (2HDMs)~\cite{Gunion:2002zf,Akeroyd:2016ymd,Branco:2011iw,Craig:2012vn}\HDMtrue\fi\xspace}
\newif\ifSM\SMfalse
\newcommand{\SM}{\ifSM SM\else standard model (SM)\SMtrue\fi\xspace}
\newif\ifMSSM\MSSMfalse
\newcommand{\MSSM}{\ifMSSM MSSM\else minimal supersymmetric standard model (MSSM)~\cite{Djouadi:2005gj,Carena:2013ytb}\MSSMtrue\fi\xspace}
\newif\ifPF\PFfalse
\newcommand{\PF}{\ifPF PF\else particle-flow (PF)~\cite{Sirunyan:2017ulk}\PFtrue\fi\xspace}
\newif\ifNLO\NLOfalse
\newcommand{\NLO}{\ifNLO NLO\else next-to-leading order (NLO)\NLOtrue\fi\xspace}
\newif\ifNNLO\NNLOfalse
\newcommand{\NNLO}{\ifNNLO NNLO\else next-to-next-to-leading order (NNLO)\NNLOtrue\fi\xspace}
\newif\ifBDTG\BDTGfalse
\newcommand{\BDTG}{\ifBDTG BDTG\else boosted decision tree with gradient boost (BDTG)\BDTGtrue\fi\xspace}
\newif\ifCL\CLfalse
\newcommand{\ConfLevel}{\ifCL \CL\else confidence level (\CL)\CLtrue\fi\xspace}
\newif\ifCSV\CSVfalse
\newcommand{\CSV}{\ifCSV CSV\else combined secondary vertex (CSV)\CSVtrue\fi\xspace}

\title{Search for charged Higgs bosons decaying into a top and a bottom quark in the all-jet final state of $\Pp\Pp$ collisions at $\sqrt{s}=13$\TeV}

\date{\today}

\abstract{
A search for charged Higgs bosons ($\PSHpm$) decaying into a top and a bottom quark in the all-jet final state is presented. The analysis uses LHC proton-proton collision data recorded with the CMS detector in 2016 at $\sqrt{s} = 13\TeV$, corresponding to an integrated luminosity of 35.9\fbinv. No significant excess is observed above the expected background. Model-independent upper limits at 95\% confidence level are set on the product of the $\PSHpm$ production cross section and branching fraction in two scenarios.  For production in association with a top quark,  limits of 21.3 to 0.007\pb are obtained for $\PSHpm$ masses in the range of 0.2 to 3\TeV. Combining this with a search in leptonic final states results in improved limits of 9.25 to 0.005\pb. The complementary $s$-channel production of an $\PSHpm$ is investigated in the mass range of 0.8 to 3\TeV and the corresponding upper limits are 4.5 to 0.023\pb. These results are interpreted using different minimal supersymmetric extensions of the standard model.
}

\hypersetup{%
pdfauthor={CMS Collaboration},%
pdftitle={Search for charged Higgs bosons decaying into a top and a bottom quark in the all-jet final state of pp collisions at 13 TeV},%
pdfsubject={CMS},%
pdfkeywords={CMS, physics, Higgs, BSM, MSSM}}

\maketitle

\section{Introduction}
\label{sec:introduction}

The observation of a
Higgs boson~\cite{Aad:2012tfa,Chatrchyan:2012xdj,Chatrchyan:2013lba,Aad:2015zhl,Sirunyan:2017exp} has motivated new areas
of study at the CERN LHC, including precision measurements of its interactions
with \SM particles~\cite{Khachatryan:2014kca,Aad:2015mxa,Khachatryan:2016vau}, searches for decays to new particles, and studies of the Higgs boson
self interactions.
Often, models beyond the \SM contain an extended Higgs sector.
Minimal extensions known as  \TwoHDMs\ include a second
complex Higgs doublet that leads to
five physical particles: two charged Higgs bosons, \PHpm, two neutral scalars,
\Ph and \PH, and one neutral pseudoscalar, \PSA.  The \TwoHDMs are further
classified according to the couplings of the doublets to fermions.  One of the popular \TwoHDMs is the \MSSM where one doublet
couples to up quarks and the other to down quarks and charged
leptons (Type-II 2HDM). In these models, the lightest CP-even Higgs boson
can align with the properties of the SM, while the additional
Higgs bosons can appear at or below the TeV scale~\cite{Carena:2013ooa}.
At tree level, the production and decay of the \PHpm\ depends on its mass (\mHpm)
and the ratio of the vacuum expectation values of the neutral
components of the two Higgs doublets ($\tanbeta$).
No fundamental charged-scalar boson is present in the \SM, and the discovery
of such a particle would uniquely point to physics beyond the \SM.

\begin{figure}[!h]
  \centering
    \includegraphics[width=0.32\textwidth]{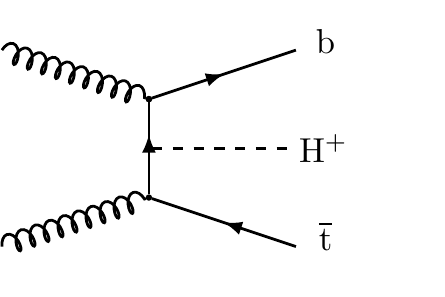}
    \includegraphics[width=0.32\textwidth]{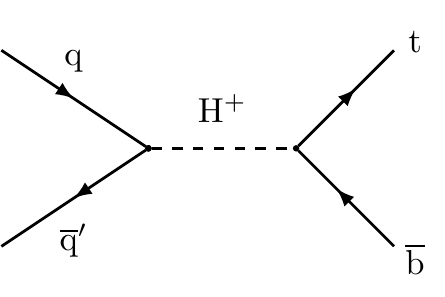}
    \caption{\label{fig:feynman}
      The LO diagrams for production of a heavy charged Higgs
      boson, showing the production with a top and a bottom quark (4FS)
      (\cmsLeft) and via an $s$-channel process (\cmsRight).
    }

\end{figure}

We report a search for charged Higgs bosons with mass larger than
that of the top quark (heavy \Hpm)
decaying to a top and bottom quark-antiquark pair (\Hptb).
The production of the boson in association with a top quark
can be described using either a four- (4FS) or a five-flavor (5FS)
scheme~\cite{deFlorian:2016spz,Harlander:2011aa},
which yield consistent results.
It can also be produced directly
via an $s$-channel process.
The corresponding leading order (LO)
Feynman diagrams are shown in \reffig{feynman}.
Charge-conjugate processes are implied throughout this paper.

Several searches for the signature \Hptb\ by the ATLAS and CMS
Collaborations in proton-proton ($\Pp\Pp$) collisions at center-of-mass energies of
8 and
13\TeV~\cite{Khachatryan:2015qxa,Aad:2015typ,Aaboud:2018cwk,Sirunyan:2019arl}
have been interpreted in the context of \TwoHDMs. Results of searches
for a light \PHpm produced in the decay $\PQt\to\PSHp\PQb$ that
subsequently decays to \csbar or \cbbar are presented in
\refcites{Khachatryan:2015uua,Sirunyan:2018dvm}.
Limits on the production of an \PHpm using the $\Pgt^{+}\Pgngt$
decay channel have also been obtained at center-of-mass energies of
8 and 13\TeV~\cite{Aad:2014kga,Khachatryan:2015qxa,Aaboud:2018gjj,Sirunyan:2019hkq}.
Charged-current processes from low-energy precision flavor
observables, such as tauonic \PB\ meson decays and the \bsgamma transition,
can be affected by the presence of the charged Higgs boson.
These results currently provide the best indirect lower limit on
\mHpm in the Type-II \TwoHDM~\cite{Bevan:2014iga,Kou:2018nap}.
Complementary searches for additional neutral heavy Higgs bosons
decaying to a pair of third generation fermions have been performed by
ATLAS and CMS at $\sqrt{s}=8$ and 13\TeV in \ttbar, $\bbbar$,
and \tautau decay
channels~\cite{Aad:2014kga,Khachatryan:2015qxa,Sirunyan:2018taj,Sirunyan:2018zut,Aaboud:2017sjh,Aaboud:2017hnm,Sirunyan:2017uyt,Sirunyan:2019wph,Aad:2019zwb}. 
The production of \PHpm via vector boson fusion with
subsequent decays to \PW and \PZ\ bosons is expected in models containing
Higgs triplets~\cite{Georgi:1985nv}. These searches are discussed in
\refcites{Aad:2015nfa,Sirunyan:2017sbn,Sirunyan:2017ret}.

The results presented here are based on $\Pp\Pp$ collision data
collected in 2016 at \sqrts{13} by the CMS experiment,
corresponding to an integrated luminosity of 35.9\fbinv.
The search investigates all-jet events, targeting a signal
containing $\PW\to\PQq\PAQq'$ decays, in the decay chains of both the
charged Higgs boson and the associated top quark.
The all-jet final state provides
the largest accessible branching fractions, $\approx$45\% and $\approx$67\%
for the top quark associated and the $s$-channel processes.
In addition, all the final state objects are detected, enabling full
reconstruction of the invariant mass of the \Hpm\ candidate.
This analysis is the first to report results in the all-jet
$\PQt\PAQb\to\PW \PQb\PAQb\to {\text{jj}}\PQb\PAQb$ channel for
top quark associated production and $s$-channel production of a charged Higgs
boson.

The search targets two distinct event topologies, boosted (for top quark associated
and $s$-channel processes) and resolved (top quark associated only).
The boosted analysis targets \Hpm bosons with mass
$\mHpm\gtrsim 5m_{\mathrm{top}}$. Decay products of \PHpm resonances with
mass of $\mathcal{O}$(TeV) have average transverse momenta (\pt) of
several hundred GeV.
As a consequence, the objects emerging from subsequent decays of
top quarks are highly collimated jets that may not be fully resolved using the
standard clustering algorithm, but can be reconstructed as a single
large-radius jet. We therefore use these collimated 
top quark- or \PW boson-jet candidates to distinguish signal events.
The resolved analysis focuses on less boosted final states where each top quark
candidate can be reconstructed from jets associated with
$\PW\to{\text{jj}}$ and one jet identified as
originating from the fragmentation of a \PQb quark (``\btagged{}").
Therefore a minimum of seven jets is expected for the associated production
channel. The search is sensitive to any narrow resonant charged state
that decays to $\PQt\PAQb$.

Model-independent upper limits on the product of the \PHpm
production cross section and branching fraction into a top and a bottom
quark ($\sigma\mathcal{B}$) as a function of \mHpm are presented
below. These limits can also be recast into model-dependent limits
and interpreted in scenario-specific limits, where the underlying
free parameters (\eg, \mHpm, branching fractions, and \tanbeta)
are fixed by the specific scenario.
Beyond the \TwoHDM interpretations, this
decay mode is relevant in the more general context of exotic resonance searches,
motivated by models of \PWpr boson production~\cite{Malkawi:1996fs,Sirunyan:2017vkm}.

\section{The CMS detector}
\label{sec:detector}

The central feature of the CMS apparatus is a superconducting solenoid
of 6\unit{m} internal diameter, providing a magnetic field of 3.8\unit{T}. Within
the solenoid volume are a silicon pixel and strip tracker, a lead
tungstate crystal electromagnetic calorimeter,  and a brass and
scintillator hadron calorimeter, each composed of a barrel and two
endcap sections. Forward calorimeters  extend  the  pseudorapidity
($\eta$) coverage beyond these barrel  and endcap
detectors.
Muons are detected in gas-ionization chambers embedded in the steel
flux-return yoke outside the solenoid.
Events of interest
are selected using a two-tiered trigger
system~\cite{Khachatryan:2016bia}. The first level, composed of
specialized hardware processors, uses information from the
calorimeters and muon detectors, while the high-level trigger consists of a
farm of processors running a version of the full event reconstruction
software optimized for fast processing. A more detailed description of
the CMS detector, together with a definition of the coordinate system
used and the relevant kinematic variables, can be found in
\refcite{Chatrchyan:2008zzk}.

\section{Event samples and simulation}
\label{sec:samples}

The main \SM backgrounds in this analysis are multijet events produced
exclusively via quantum chromodynamics (QCD) interactions
and top quark-antiquark pair production. Other sources of background
include single top quark production and $\PQt\PAQt{+}\PX$ processes with
\PX = (\PW, \PZ, \Pgg, \PH, or \ttbar), and also \PV{+}jets
(\PV = \PZ\ or \PW), diboson ($\PW\PZ$, $\PZ\PZ$, $\PW\PW$, $\PV\PH$)
and triboson processes.
The latter group is denoted as the ``Electroweak'' background below.

Simulated samples are produced using various \MC event generators.
Signal samples are generated using the 4FS at \NLO precision in
perturbative QCD with the \MGvATNLO~v2.3.3~\cite{Alwall:2014hca}
generator for a range of \mHpm hypotheses from \massRangeAssCh.
The total cross section for
the \PHpm production associated with a top quark is obtained using the
\textit{Santander matching} scheme~\cite{Harlander:2011aa}.
Typical values are of the order of 1\pb\ for $\mHpm=0.2\TeV$, down to
about $10^{-4}$\pb\ for a mass of 3\TeV~\cite{Heinemeyer:1998yj,deFlorian:2016spz,Berger:2003sm,Flechl:2014wfa,Degrande:2015vpa,Dittmaier:2009np}.
The $s$-channel signal processes are simulated using LO
\textsc{CompHEP}~4.5.2~\cite{Boos:2004kh} following a $\PWpr_{\text{R}}$ model
in the narrow-width approximation, in the mass range
from 0.8 to 3\TeV~\cite{Sirunyan:2017vkm}.
Branching fractions \BHptb\ are computed using
\textsc{hdecay}~v6.25~\cite{Djouadi:1997yw} for different values of \tanbeta.

Both the QCD multijet and  \PV{+}jets background samples are simulated at LO using the \MGvATNLO~v2.2.2 event generator.
The \ttbar sample is generated using
\POWHEG~v2~\cite{Nason:2004rx,Frixione:2007vw,Alioli:2010xd}
at \NLO in QCD~\cite{Frixione:2007nw}, assuming a \TopQuark mass of 172.5\GeV.
Single \TopQuark events are generated at \NLO precision in the 4FS
for the $t$-channel process~\cite{Frederix:2012dh} using \POWHEG~v2
interfaced with \textsc{madspin}~\cite{Artoisenet:2012st} for simulating
the \TopQuark decay.  The $s$-channel process is simulated using
\MGvATNLO~v2.2.2, while the production of single top quark events via
the $\PQt\PW$ channel is simulated at NLO in the 5FS using
\POWHEG~v1~\cite{Re:2010bp}. The ``diagram removal" approach~\cite{Frixione:2008yi}
is used to avoid the partial double counting of $\PQt\PW$ production vs. \ttbar production at NLO.
The production of \ttbar in association
with \PW, \PZ, or \Pgg is simulated at NLO using \MGvATNLO~v2.2.2.
The production of $\ttbar\PH$, where \PH decays to a $\bbbar$ pair is
generated using \POWHEG~v2 at \NLO~\cite{Hartanto:2015uka}.
The samples are normalized to the most precise available cross
section calculations, corresponding most often to
\NNLO in QCD and \NLO in electroweak
corrections~\cite{Kidonakis:2012db,Cacciari:2011hy,Baernreuther:2012ws,Czakon:2012zr,Czakon:2012pz,Beneke:2011mq,Czakon:2013goa,Czakon:2011xx,Aliev:2010zk,Kant:2014oha,Maltoni:2015ena,Campbell:2011bn,Kidonakis:2010ux}.

Parton distribution functions (PDFs) are modeled using the
NNPDF3.0~\cite{Ball:2011uy} parametrization. Parton showering and
fragmentation are performed using the \PYTHIA~v8.212~\cite{Sjostrand:2014zea}
generator.  The CUETP8M2T4~\cite{Sirunyan:2018avv}  tune is used to characterize
the underlying event in the \ttbar background, while the
CUETP8M1~\cite{Skands:2014pea,Khachatryan:2015pea} tune is used for all
other processes.

The response of the CMS detector for all generated samples
is simulated using \GEANTfour~v9.4~\cite{Agostinelli:2002hh}.
Additional $\Pp\Pp$ interactions in the same or nearby bunch
crossings (pileup) are generated with \PYTHIA~v8.212 and superimposed on the
hard collisions.  In the data collected in 2016, an average of 23
$\Pp\Pp$~interactions occurred per LHC bunch crossing.
Simulated events are corrected to produce the pileup distribution in data
based on the measured luminosity profile and average measured total inelastic
$\Pp\Pp$ cross section~\cite{Sirunyan:2018nqx}.

\section{Object reconstruction}
\label{sec:obj}

Events are processed using the \PF algorithm, which aims to
reconstruct and identify all particles (\PF candidates) using the optimal combination
of information from the tracker, calorimeters, and muon systems of the
CMS detector.

Electron candidates are identified by matching clusters of
energy deposits in the electromagnetic calorimeter to reconstructed
charged-particle trajectories in the tracker.
A number of selection criteria based on the shower shape,
track-cluster matching, and consistency between the cluster energy and
track momentum are then applied for the identification of
electrons~\cite{Khachatryan:2015hwa}.
Muons are reconstructed by requiring consistent hit patterns in the tracker and
muon systems~\cite{Sirunyan:2018fpa}.
The relative isolation variable for an electron or muon candidate
is defined as the scalar sum of the transverse momenta of
all \PF candidates in a cone around
the candidates' trajectory divided by the lepton \pt.
The cone size depends on the lepton \pt, and is bounded between a distance parameter of 0.05 and of 0.2.
Hadronically decaying $\tau$ leptons of $\pt\ge~20\GeV$, within
$\absEta=2.3$ are reconstructed using the hadron-plus-strip
algorithm~\cite{Sirunyan:2018pgf}. The corresponding isolation variable
is computed using a multivariate approach, combining information on
its identification, isolation and lifetime~\cite{Sirunyan:2018pgf,Khachatryan:2015dfa}.

The primary jet collection is formed by clustering \PF candidates using
the anti-\kt algorithm~\cite{Cacciari:2008gp, Cacciari:2011ma} with a
distance parameter of 0.4.  Jet energies are corrected for contributions coming
from event pileup~\cite{CMS:2019kuk}. Additional corrections to the jet energy
scale~\cite{Khachatryan:2016kdb}
are applied to compensate for nonuniform detector response. Jets are
required to have $\pt>40\GeV$ and be contained within the tracker
volume of $\absEta<2.4$.  For the resolved analysis, the \pt requirement
is relaxed to 30\GeV for subleading jets ranking seventh or lower in \pt.
In the boosted analysis an additional large-radius jet collection is defined
using a distance parameter of 0.8.

Jets consistent with originating from the decay of a \PQb quark are identified
using the \CSV \btagging algorithm~\cite{Sirunyan:2017ezt}, at the medium or
loose working points. These are defined such that the efficiency to select
light-flavor quarks (\PQu, \PQd, or \PQs) or gluons as \PQb jets
is about 1\% (medium) or 10\% (loose), and the corresponding efficiency for tagging jets
from a \PQb(\PQc)\ quark decay is about 65 or 80\% (10 or 25\%), respectively.
For brevity
we refer to jets satisfying the \btagging criteria as \PQb jets below.

The scalar \pt sum of all selected jets in an event is denoted as \HT,
while the missing transverse momentum vector \ptvecmiss
is defined as the projection onto the plane perpendicular to the
beam axis of the negative vector sum of the momenta of all
reconstructed \PF candidates~\cite{Sirunyan:2019kia}. Its magnitude
is referred to as \ptmiss. Quality requirements are applied to remove a small
fraction of events in which detector effects, such as the electronic noise,
can affect the \ptmiss reconstruction~\cite{Sirunyan:2019kia}.
The energy scale and resolution
corrections applied to jets are propagated to the calculation
of \HT and \ptmiss.

\section{Search strategy}
\label{sec:strategy}

The analysis aims to reconstruct the full event in order to
search for a local enhancement in the top and bottom quark-antiquark
invariant mass spectrum.

Because of the large cross section for the QCD multijet background, restrictive
trigger requirements are needed to reduce the data recording rate. The data
used for this search are collected with an inclusive online selection of
$\HTtrig>900\GeV$, with \HTtrig being defined as the scalar \pt\ sum of
small-radius jets with $\pTtrig > 30\GeV$. Events are also acquired with a
dedicated
large-radius jet trigger requiring $\pTtrig > 360\GeV$ and a mass
after jet trimming~\cite{Krohn:2009th} of at least 30\GeV.
Furthermore, events satisfying trigger requirements of
$\HTtrig > 450\,(400)\GeV$ and six jets with $\pTtrig > 40~(30)\GeV$
are selected if at least one (two) of them satisfies \btagging criteria.
In the low \mHpm regions the sensitivity of the all-jet final state is
limited by the relatively high trigger thresholds.

Two analyses are performed, each targeting different regions of the signal
parameter space. The \boostedAnalysis\ targets charged Higgs bosons with
high mass and utilizes collimated hadronically decaying
\PW boson or \TopQuark candidates to distinguish signal events.
A collection of large-radius jets is used to reconstruct and identify
the objects from the decays of boosted \PW boson and top quark.
In order to discriminate against QCD multijet backgrounds, we exploit both
the reconstructed jet mass, which is required to be close to the
\PW boson or \TopQuark mass, and the two- or three-prong
jet substructure (subjets) corresponding to the $\PW\to\PQq\PAQq'$ or
$\PQt\to\PQq\PAQq'\PQb$ decay~\cite{CMS-PAS-JME-16-003}.
The soft-drop algorithm~\cite{Larkoski:2014wba} is used to remove soft
and wide-angle radiation.  The use of soft-drop grooming reduces the resulting
jet mass \mSD\ for QCD multijet events where large jet masses arise
from soft-gluon radiation.
Finally, because top-quark jets contain a \PQb quark and \PW jets do not,
additional discrimination power is achieved by applying the CSV algorithm
described above to the constituent subjets. The events are categorized
according to the number of \btagged jets to separate
sources of \SM background and capture signals with both high and low
number of \PQb quarks.

The  \resolvedAnalysis\ is optimized for charged Higgs bosons with lower masses
that decay to  moderately boosted top quarks, often identified as three
separate small-radius jets, one of which is \PQb tagged and the other two
jets resulting from the \PW boson decay.  The resolved top quark
candidates (\toptaggingRes) are identified using a multivariate \BDTG
classifier. The classifier exploits properties of the
top quark and its decay products such as masses, angular separations,
and other kinematic distributions. Additional input variables are
quark vs. gluon~\cite{CMS:2013kfa}, charm vs. light quark~\cite{CMS:2016knj},
and \btagging discriminator values for each of the three jets.
The signal enriched region is defined by requiring the presence of
seven or more jets, comprising two resolved top quark candidates
and an additional \btagged jet used to reconstruct the \PHpm candidate.

Both analyses veto the presence of an isolated charged lepton
(\Pe or \Pgm) with $\pt\ge 10\GeV$, or an isolated hadronically
decaying tau lepton with $\pt\ge 20\GeV$.
The lepton veto ensures that
leptonic final states of \PW bosons produced in top quark decays are not
considered. These are covered by dedicated analyses~\cite{Sirunyan:2019arl}.
To further reduce the background from semileptonic \ttbar decays,
the \boostedAnalysis requires events to have
$\ptmiss<200\GeV$. These requirements also reduce background from any sources
containing \PW and \PZ boson decays.

\subsection{Event categories in the boosted analysis}

Events with at least one \btagged jet and one identified \TopQuark
candidate are considered in the boosted analysis. The \PHpm candidate
four-momentum vector is reconstructed as the sum of the four-vectors of
the loose-tagged \bjet with highest \pt and the top quark candidate with mass most
closely matching $m_{\text{top}}$. A top quark candidate is identified as
a \tjet or the combination of a \Wjet and a \bjet, excluding the \bjet with
highest \pt.
The {\PW}(t) jet candidates are required to have 
$65<\mSD<105\GeV$ ($135<\mSD<220\GeV$), $\pt>200~(400)\GeV$,
and $\absEta<2.4$. The hard substructures are identified using the
$N$-subjettiness~\cite{Thaler:2010tr} ratios:
${\tau_{2}}/{\tau_{1}}<0.6$ for the \PW jet and
${\tau_{3}}/{\tau_{2}}<0.67$ for the \tjet.
We introduce four mutually exclusive categories. The labels
``t1b'' and ``t0b'' refer to events containing a large-radius jet identified
as a \tjet, where at least one, or exactly zero, of the
subjets satisfies the medium working point of the \PQb tagging algorithm.
In the ``wbb'' category the top quark candidate is formed from a \Wjet and
an additional medium-tagged \PQb jet.  The ``wbj'' category
relaxes the \btagging requirement on one of the additional jets to satisfy the
loose \btagging working point.

\begin{figure}[t]
\centering
  \includegraphics[width=0.80\textwidth]{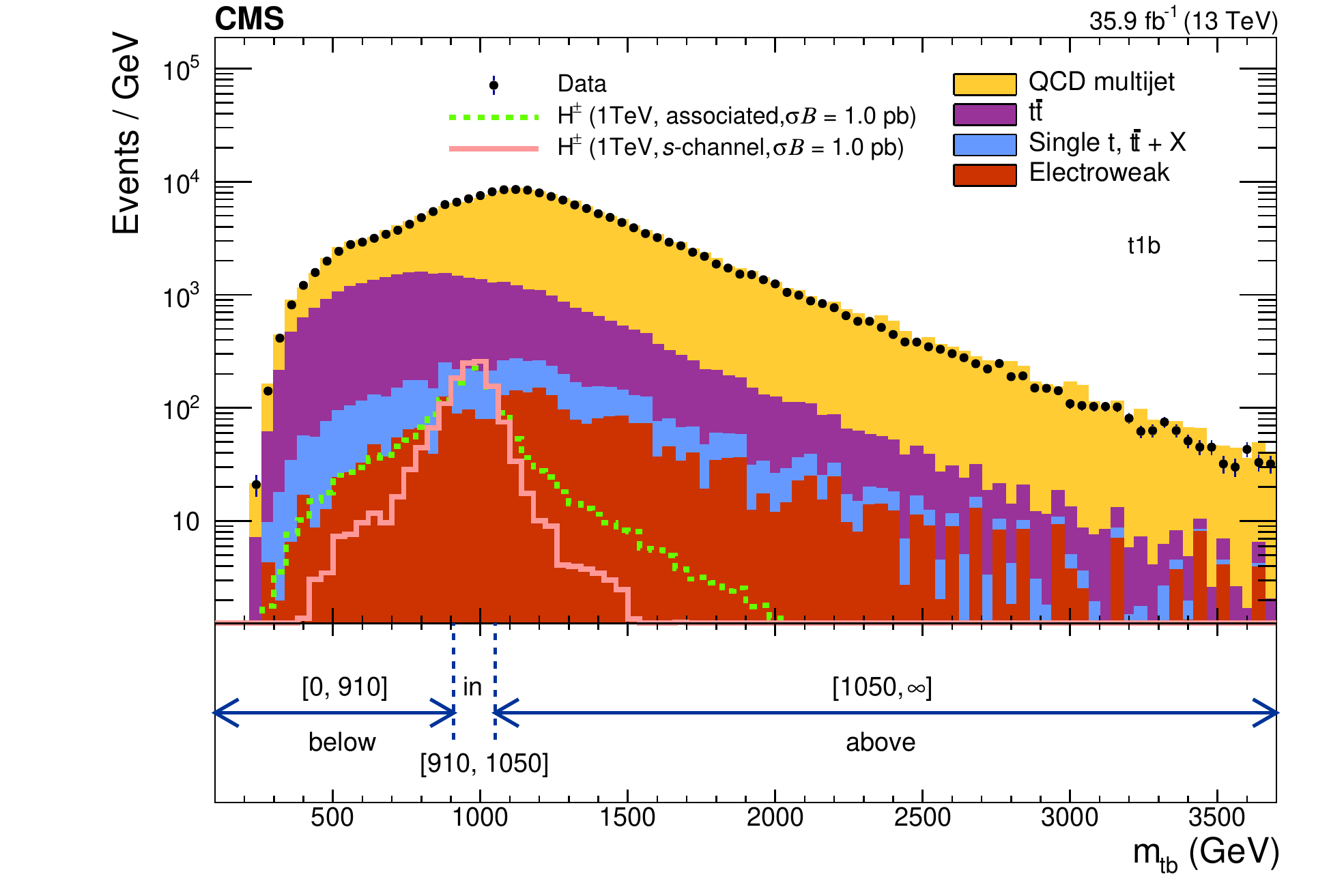}
  \caption{\label{fig:MassCandidate_Signal}
    Data and \SM background for the event sample with one \tjet as a function of
    the charged Higgs boson candidate mass. The category t1b is shown and
    the background normalization is fixed to the SM expectation.
    The signal mass distributions for associated and $s$-channel production
    of an \Hpm with $\mHpm=1\TeV$ normalized with a cross section times branching
    fraction of 1\pb are superimposed as open histograms.
    The signal mass window ``in'' for associated production
    is shown together with the sidebands ``below'' and ``above'' for the
    mass hypothesis of 1\TeV.
  }

\end{figure}

The signal is characterized as a peak in the invariant mass
distribution $m_{\PQt\PQb}$ of \Hpm candidates. This distribution is dominated
by background contributions from QCD multijet processes.
The expected shape of the \Hpm candidate mass distribution
is dominated by the detector resolution and pairing errors,
where jets are not correctly matched to the decay products of the boson;
the latter is primarily responsible for the sideband on the left of the peak
of the invariant mass distribution.
The full width at half maximum of the reconstructed MC mass
distribution for correct jet assignments is used to describe the mass
resolution, and events falling outside this window are used to constrain
the background.
The mass resolution is consistent among the different event categories. For
a charged Higgs boson of mass 1\TeV the resolution is approximately 140\GeV.  
The distribution of the invariant mass $m_{\PQt\PQb}$ for the t1b category
is shown in Fig.~\ref{fig:MassCandidate_Signal} for the data,
the expected background and for a signal of mass $\mHpm=1\TeV$.
To enhance the expected signal to background ratio, data are selected
within a window around different charged Higgs boson candidate masses
in each of the categories listed above.
We then search for an excess of events in the \HT data distribution.

To better separate signal from background, the event categories are
further subdivided
to exploit differences in jet and \bjet multiplicities. For signal events
produced in association with a top quark, we expect at least three
\mbox{\PQb quarks} in the final state and a large number of extra jets
not participating in the reconstruction of the \PHpm. Signal produced
in the $s$~channel contains two \PQb quarks and fewer extra jets.
We therefore consider different requirements on the number of
\btagged jets: exactly one, exactly two, and at least three.
We also distinguish two categories based on the number of
additional small-radius jets,
less than three ($N_{\text{jets}}<3$) or at least three ($N_{\text{jets}} \ge 3$) such jets.

The signal-rich regions are analyzed together with signal-depleted
regions using a binned maximum likelihood fit to the \HT data distributions that simultaneously
determines the contributions from signal and the major
background sources.

\subsection{Event selection in the resolved analysis}
A multivariate analysis is employed to select
top quark candidates in events containing seven or more jets.
We employ a \BDTG classifier that is 
trained on simulated \TopAntitopQuark pair events using
the \textsc{tmva} package~\cite{TMVA}.
The signal objects are considered to
be three small-radius jet combinations, in which each individual jet is
matched to the decay product of a top quark at the generator
level. Similarly, background objects are defined as three-jet
combinations in which at least one jet is not matched to a
\TopQuark decay product. The input variables used for the
\BDTG training (19 in total), calculated from these jet combinations,
are described in detail in \refcite{Sirunyan:2017wif}. In the \BDTG
response distribution, values close to $-1$ are mainly
populated by fake \TopQuark candidates from QCD multijet processes, while
values close to $+1$ are dominated by \TopQuark candidates from \ttbar or
signal events.  In this analysis we require resolved \TopQuark candidates
to have a BDTG score $> 0.4$, yielding a signal object efficiency of
92\%, and a background object efficiency of 6\%.

Events with at least three \btagged jets passing the CSV medium working point
and at least four additional jets are selected.  The first \TopQuark candidate
is identified by pairing each \btagged jet with all two-jet combinations and
retaining the combination with the highest BDTG value.  The same procedure is
applied for the second candidate, using only the remaining jets as inputs.
To reduce the combinatorial background, we require
all the combined three-jet systems to have
an invariant mass less than 400\GeV.

\begin{figure}[t]
\centering
  \includegraphics[width=0.49\textwidth]{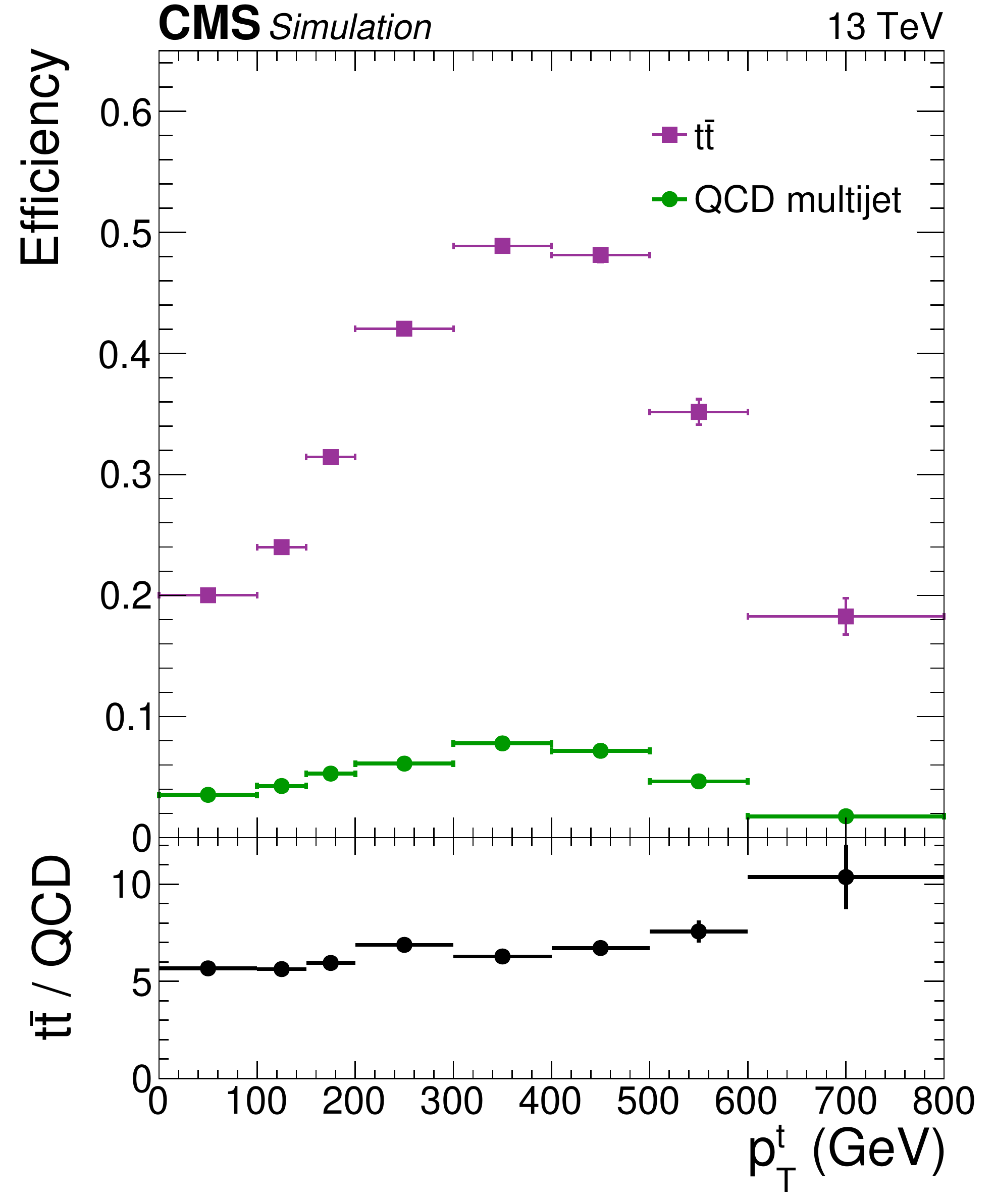}
  \includegraphics[width=0.49\textwidth]{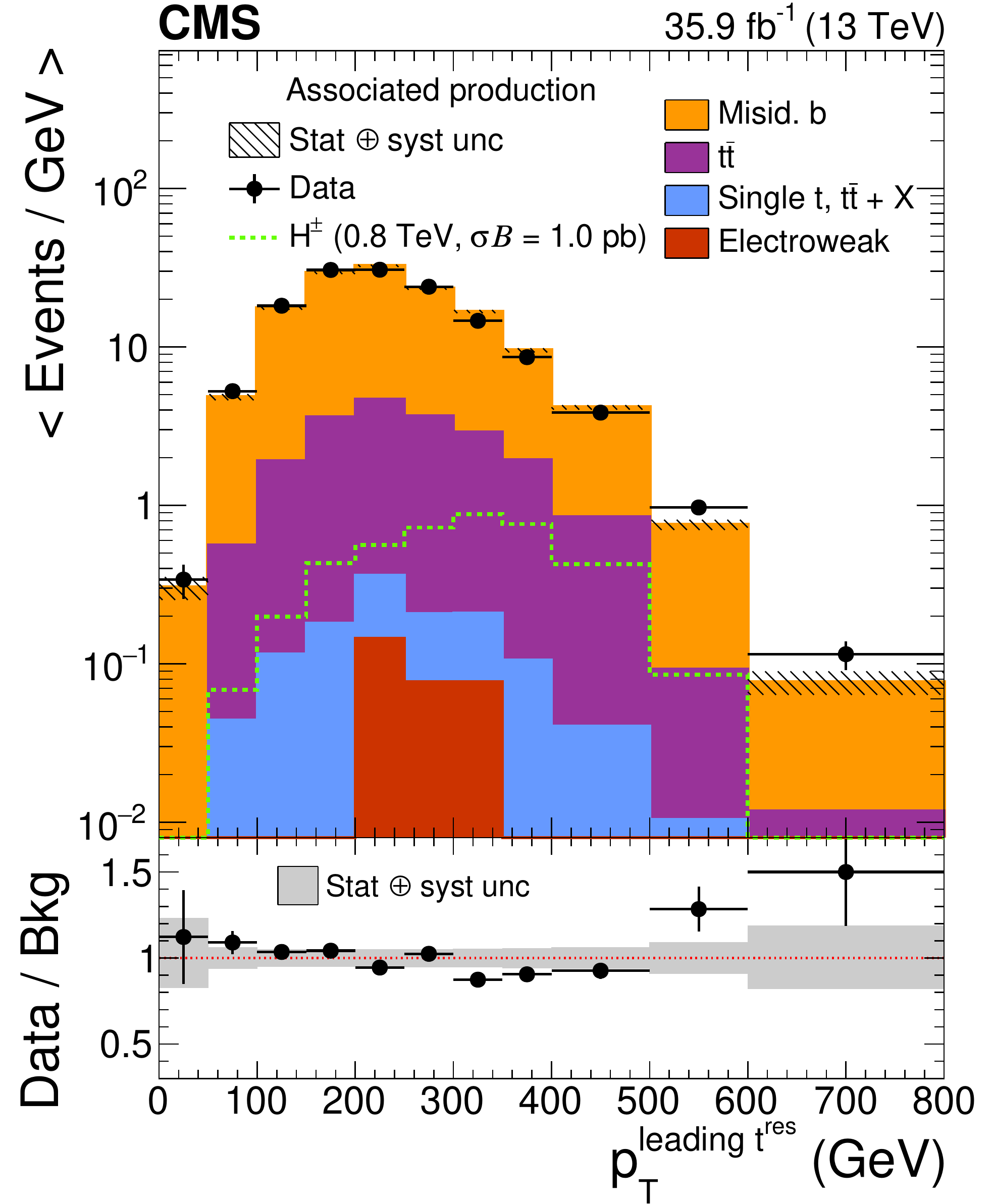}
  \caption{\label{fig:resolved-top}
    The efficiency of the \toptaggingRes\ selection in simulated
    \ttbar pairs and the misidentification rate for QCD multijet background,
    as a function of top quark or top quark candidate \pt,
    respectively (\cmsLeft).  The \pt distribution of the leading
    \toptaggingRes (\cmsRight) for the signal model and background with
    normalization fixed to the SM expectation. The dominant background
    containing misidentified {\bjet}s is primarily composed of QCD
    multijet processes and is estimated using a data-driven method.
    The expectation for a signal with $\mHpm=0.8\TeV$ is also shown.
  }

\end{figure}

The efficiency of the BDTG requirement as a function of the \pt of
the generated top quark in \ttbar events is shown in
\reffig{resolved-top} (\cmsLeft), along with the
misidentification rate observed in a QCD multijet sample. At the
plateau the tagging efficiency reaches 50\%. The observed decrease
in efficiency in the high-\pt region is due to \TopQuark decay products
becoming increasingly collimated, resulting in a jet-to-parton
matching inefficiency. The misidentification rate is less than 8\%
for the entire \pt range considered.

To reconstruct the invariant mass of the \PHpm candidate,
we use the resolved top quark candidate with the highest
transverse momentum, $\pt^{\text{leading}~\toptaggingRes}$,
and the \btagged jet having highest \pt that is not used in the
reconstruction of the two selected \TopQuark candidates.
The distribution of $\pt^{\text{leading}~\toptaggingRes}$ is
shown in \reffig{resolved-top} (\cmsRight).
The invariant mass $m_{\PQt\PQb}$ of the \PHpm candidate is used in a binned
maximum likelihood fit to extract the signal in the presence of the \SM
background.

\section{Backgrounds}
\label{sec:bkg}

The dominant backgrounds arise from QCD multijet processes and
\TopQuark pair production in association with additional jets.
Contributions from more rare processes, such as
single top quark, $\PQt\PAQt{+}\PX$, \PV{+}jets, diboson, and triboson production
are found to be small.

\subsection{Background estimations in events with boosted \texorpdfstring{\PW}{W} boson and top quark candidates}
We estimate the QCD multijet and \TopQuark backgrounds using a method
that exploits a number of background-rich control regions (CRs) in data.
These control regions are included in a simultaneous fit with the signal
enriched regions to determine the normalization and the shape of the
background distributions.

Because the cross section for QCD multijet production is large, this background
can produce many events satisfying the signal selection requirements.
The distribution of \mSD for signal peaks around the \PW boson or
the \TopQuark mass for large-radius jets corresponding to their hadronic
decays, while the QCD multijet background spectrum is peaked at lower \mSD.
This background is estimated from simulation with corrections applied to both
shape and normalization.  These corrections are determined by matching the
simulation to data using a CR enriched with jets arising from the
hadronization of single quarks or gluons.  The CR is defined by inverting
the $N$-subjettiness requirements used to identify the \PQt and \PW jets.
The shape is determined for each event category using both this CR
and the sideband regions around the signal mass windows in the invariant
mass spectrum of the top and bottom quark pair, while the normalization is
determined from the sidebands only.  We validate this correction by applying
the technique in an orthogonal CR defined by requiring that no \btagged
jets are identified. The shapes of the $N$-subjettiness distributions and
kinematics of jets having \mSD\ consistent with either a \PQt or \Wjet
are found to be consistent with events passing the signal selection.

The contribution from the \ttbar process arises from all-jet
final states or with a leptonic decay of a \PW boson where the charged
lepton is outside the kinematic acceptance of the CMS detector or evades
identification by the dedicated lepton vetoes. Such events contain a
pair of \PQb quarks and boosted \PW and {\tjet}s.
The \ttbar background is estimated from simulation and
normalized using a CR in data.
A lepton enriched set of events is
used to describe the kinematics for the top quark pair
production and the normalization is allowed to vary unconstrained in the
final fit. The CR is defined by requiring a lepton (\Pe, \Pgm) with
$10<\pt<35\GeV$, $\ptmiss>100\GeV$ and at least one \btagged jet.
This ensures orthogonality with the searches for charged Higgs
bosons in the leptonic channels~\cite{Sirunyan:2019arl}.

\subsection{Background estimations in events with resolved top quarks}
The main backgrounds for the \resolvedAnalysis
can be decomposed into events containing either genuine \PQb jets
or events with at least one light quark or gluon jet erroneously tagged
as a \bjet.  We refer to the latter as misidentified {\bjet}s.
The background containing genuine \PQb jets is modeled using simulation.
The background due to misidentified {\bjet}s is measured with a
data-driven technique using control regions that are defined
by inverting the BDTG requirement, the \PQb jet selection, or both.

The shape of the \PHpm candidate mass distribution in the
background is obtained from events that are separated from the signal
region (SR) by requiring that only two (of at least three) \PQb jets pass
the \CSV medium working point, and the remaining jets only pass the loose \CSV
working point. This region is referred to as the application region (AR). In
order to compensate for the different selection efficiencies between
these two regions, transfer factors are used to normalize
the AR to the SR. These transfer factors are determined by taking the
ratio of events in two additional CRs that are orthogonal
to each other and to both the AR and SR.
The first CR, CR1, is obtained by requiring one \toptaggingRes\ candidate
plus a second top quark candidate failing the BDTG requirement,
and the second CR, CR2, is obtained by
also altering the \PQb jet selection as described above for the AR.
For the regions defined as AR and CRs, a correction is applied to
remove events containing jets from b quark decays that fail the tagging
requirement.
In order to minimize the effect of kinematic differences between
the loose and medium working points, the background from misidentified
{\bjet}s is evaluated separately in \pt and $\eta$ bins of the
\btagged jet used in the reconstruction of the invariant mass of the
\PHpm candidate.

Because the SR and associated CRs are mutually exclusive,
the expected yield of misidentified \bjet events passing the signal selections
can be predicted as:
\begin{linenomath}
\begin{equation}
\label{eq:Fakeb}
N^{\text{SR}} =
\sum\limits_{i} N_{i}^{\text{AR}}
\left(\frac{N_{i}^{\text{CR1}}}{N_{i}^{\text{CR2}}}\right)\!,
\end{equation}
\end{linenomath}
where CR1(2) refers to the first (second) control region and the
index $i$ runs over all \pt and $\eta$ bins of the aforementioned
highest \pt \btagged jet.

\section{Systematic uncertainties}
\label{sec:systematics}

The systematic uncertainties are divided into two categories: those
that affect the estimation of the background from the \SM processes, and
those that affect the expected signal distributions and yields.

The events used in this search are largely collected with a trigger
efficiency close to 100\%. The trigger efficiency is extracted from
data and the uncertainties in trigger correction factors
applied to the simulation are less than 5\%.

The uncertainty from pileup modeling is estimated by varying the total
inelastic $\Pp\Pp$ cross section of 69.2 mb by 5\%~\cite{Aaboud:2016mmw}.
The uncertainty in the integrated luminosity is estimated to be
2.5\%~\cite{CMS:2017sdi}.

Uncertainties in the background prediction that also affect the signal
arise from the jet energy scale~\cite{CMS-PAS-JME-16-003}, from the scale
factors correcting the efficiency and misidentification rate for
\btagging~\cite{Sirunyan:2017ezt}, and from the reconstruction and
identification efficiencies of the leptons. In addition, uncertainties
arising from the simulation-to-data corrections for boosted \PQt and
\PW tagging and the BDTG response are applied in the boosted and resolved
analyses, respectively.
The variations in the jet selection and jet energy scale are propagated to
the \HT, \ptmiss, and $\PSHpm$ candidate yields and invariant mass.

For the \boostedAnalysis a normalization uncertainty of 50\% is applied
for the QCD multijet background.  This uncertainty is treated as
uncorrelated among the boosted \PQt- and \PW-tagged event categories and
it is 100\% correlated within each category and across the signal regions
and the sidebands.
An additional uncertainty to account for shape variations in  modeling
the \HT observable is parametrized linearly as a function of \HT and
reaches 30\% for an \HT of 1\TeV.
These uncertainties are then constrained by studying the CR used to
correct the simulation and the resulting variation in expected
QCD multijet background yield is approximately 28\%.

The systematic uncertainties affecting the misidentified \bjet background
measurement in the \resolvedAnalysis can be divided into three
components. The first component consists of events containing jets from
\PQb quark decays that fail the \btagging requirement and is subtracted from
the CRs used in the measurement. The uncertainty on the normalization
of this component is estimated by propagating all the uncertainties related
to the simulation of electroweak and top quark processes.
The other two components account for the statistical and systematic
uncertainties in determining the transfer factors applied in the normalization
of the AR. Statistical fluctuations in the value of the transfer factors can
result in rate and shape differences in the predicted invariant mass
distribution. Similarly, the definition of the CR affects
the individual transfer factors and subsequently the invariant mass shape
in the AR. The aforementioned contributions affect the expected event
yield by approximately 4\%.

For \ttbar and single \TopQuark processes, the effect of the \TopQuark mass on
the cross sections is estimated by varying the \TopQuark mass by $\pm 1.0\GeV$
around the nominal value of 172.5\GeV.

Theoretical uncertainties in both the acceptance and the cross sections are determined
by varying the choice of factorization and renormalization scales and PDFs.
Uncertainties due to scales in the inclusive cross sections are estimated for each simulated process by
varying the scales independently and together by factors of
0.5 and 2 with respect to the default values. The event yields are then
calculated for each of the six variations and the maximum variation
with respect to nominal is taken as the systematic uncertainty.
The PDF uncertainties are treated as fully correlated for all processes
that share the same dominant partons in the initial state of the matrix
element calculations (\ie, $\Pg\Pg$, $\Pg\PQq$, or $\PQq\PQq$)~\cite{Butterworth:2015oua}.

Finally, the limited numbers of simulated background and signal events
lead to statistical fluctuations in the nominal predictions.
The effects are considered in the limit calculations using a
\textit{Barlow--Beeston lite} approach~\cite{Barlow:1993dm,Conway:2011in},
which assigns the combined statistical uncertainty in each bin to the
overall background yield in that bin.

Tables~\ref{tab:Syst_Boosted} and~\ref{tab:systematics-resolved}
summarize the various sources of systematic uncertainty and their
impact on the signal yield and the total expected background in data, 
for the boosted and resolved analyses, respectively.

\begin{table}[!h]
\centering
\topcaption{
The systematic uncertainties affecting signal and background for the
\boostedAnalysis, evaluated after fitting to data, summed over all
final states and categories.  The numbers are given in percentage and
describe the effect of each nuisance parameter on the overall normalization
of the signal model or the total background.
Nuisance parameters with a check mark also affect the
shape of the \HT spectrum.  Sources that do not apply in a given process
are marked with dashes.
For the $\PSHpm$ signal, the values for $\mHpm=1\TeV$ and for
associated production are shown.}
\label{tab:Syst_Boosted}
\renewcommand{\arraystretch}{1.0} 
\cmsTable{
\begin{tabular}{ l c c c c c c c}

Source 	&	 Shape  	&	 $\PSHpm$ 	&	 Multijets 	&	 \ttbar 	&	 Single \PQt, $\ttbar{+}\PX$ 	&	 Electroweak	\\
\hline													
Trigger efficiency 	&	 	&	5.0	&	4.5	&	0.39	&	0.05	&	0.06	\\
Pileup 	&	 $\checkmark$ 	&	0.42	&	1.4	&	0.05	&	 $<$0.01 	&	0.03	\\
Integrated luminosity 	&	 	&	2.5	&	 \mc{\NA} 	&	0.2	&	0.02	&	0.03	\\
Lepton efficiency 	&	 	&	5	&	 \mc{\NA} 	&	0.39	&	0.05	&	0.06	\\
Jet energy scale and resolution 	&	 $\checkmark$ 	&	3.0	&	5.8	&	0.4	&	0.04	&	0.12	\\
\PQb jet identification 	&	 $\checkmark$ 	&	2.4	&	12	&	0.24	&	0.03	&	0.12	\\
Unclustered \ptmiss energy scale 	&	 $\checkmark$ 	&	0.23	&	 \mc{\NA} 	&	0.02	&	 $<$0.01 	&	0.01	\\
Jet $m_{\text{SD}}$ scale 	&	 $\checkmark$ 	&	1.3	&	2.5	&	0.07	&	0.02	&	 \mc{\NA} 	\\
$N$-subjettiness scale 	&	 	&	2.0	&	 \mc{\NA} 	&	0.17	&	0.02	&	 \mc{\NA} 	\\
QCD bkg. normalization 	&	 	&	 \mc{\NA} 	&	28	&	 \mc{\NA} 	&	 \mc{\NA} 	&	 \mc{\NA} 	\\
QCD bkg. shape 	&	 $\checkmark$ 	&	 \mc{\NA} 	&	 $<$0.01 	&	 \mc{\NA} 	&	 \mc{\NA} 	&	 \mc{\NA} 	\\
Top quark mass 	&	 	&	 \mc{\NA} 	&	 \mc{\NA} 	&	0.21	&	0.02	&	 \mc{\NA} 	\\
{}                     	&	              	&	             	&	              	&	      	&		&		\\
Theory source                     	&	              	&	             	&	              	&	      	&		&		\\ \hline
Scales, PDF (acceptance) 	&	 $\checkmark$ 	&	2.1	&	 \mc{\NA} 	&	0.53	&	 \mc{\NA} 	&	 0.04 	\\
Scales, PDF (cross section) 	&	 	&	 \mc{\NA} 	&	 \mc{\NA} 	&	0.43	&	0.04	&	 0.05 	\\
\end{tabular}
}
\end{table}

\begin{table}[!h]
\centering
\topcaption{
The systematic uncertainties in the backgrounds and the signal for the
\resolvedAnalysis, evaluated after fitting to data. The numbers are given in percentage and
describe the effect of each nuisance parameter on the overall normalization
of the signal model or the total background.
Nuisance parameters with a check mark also affect the
shape of the $\PSHpm$ candidate mass spectrum.  Sources that do not apply
in a given process are marked with dashes.
For the $\PSHpm$ signal, the values for $\mHpm=0.5\TeV$ are shown.}
\label{tab:systematics-resolved}
\renewcommand{\arraystretch}{1.0} 
\cmsTable{
\begin{tabular}{ l c c c c c c c}

Source 	&	 Shape  	&	 $\PSHpm$ 	&	 Misid. \PQb 	&	 \ttbar 	&	 Single \PQt, $\ttbar{+}\PX$ 	&	 Electroweak	\\
\hline													
Trigger efficiency     	&	              	&	5.0	&	0.09	&	0.69	&	0.04	&	0.01	\\
Pileup          	&	 $\checkmark$ 	&	 $<$0.01 	&	  \mc{\NA}      	&	 $<$0.01 	&	 $<$0.01  	&	 $<$0.01	\\
Integrated luminosity             	&	              	&	2.5	&	0.09	&	0.35	&	0.02	&	 $<$0.01	\\
Lepton efficiency 	&	              	&	0.32	&	  \mc{\NA}      	&	0.04	&	 $<$0.01 	&	 $<$0.01	\\
Jet energy scale and resolution 	&	 $\checkmark$ 	&	8.5	&	0.24	&	1.6	&	0.09	&	0.33	\\
\PQb jet identification 	&	 $\checkmark$ 	&	5.0	&	 \mc{\NA} 	&	0.64	&	0.04	&	0.01	\\
\toptaggingRes tagging         	&	 $\checkmark$ 	&	8.9	&	 0.24     	&	1.8	&	0.04	&	 $<$0.01	\\
Transfer factors           	&	 $\checkmark$ 	&	 \mc{\NA}    	&	4.0	&	 \mc{\NA} 	&	 \mc{\NA}    	&	 \mc{\NA}	\\
Top quark mass      	&	     	&	 \mc{\NA} 	&	0.09	&	0.39	&	0.02	&	 \mc{\NA} 	\\
{}                     	&	              	&	             	&	              	&	      	&		&		 \\
Theory source                     	&	              	&	             	&	              	&	      	&		&		 \\ \hline
Scales, PDF (acceptance) 	&	  	&	5.1	&	 \mc{\NA} 	&	0.39	&	0.02	&	0.01	\\
Scales, PDF (cross section) 	&	     	&	 \mc{\NA} 	&	0.12	&	0.76	&	0.04	&	0.01	\\
\end{tabular}
}
\end{table}

\section{Results and interpretation}
\label{sec:results}

\begin{figure}[thb]
\centering
\includegraphics[width=1.0\textwidth]{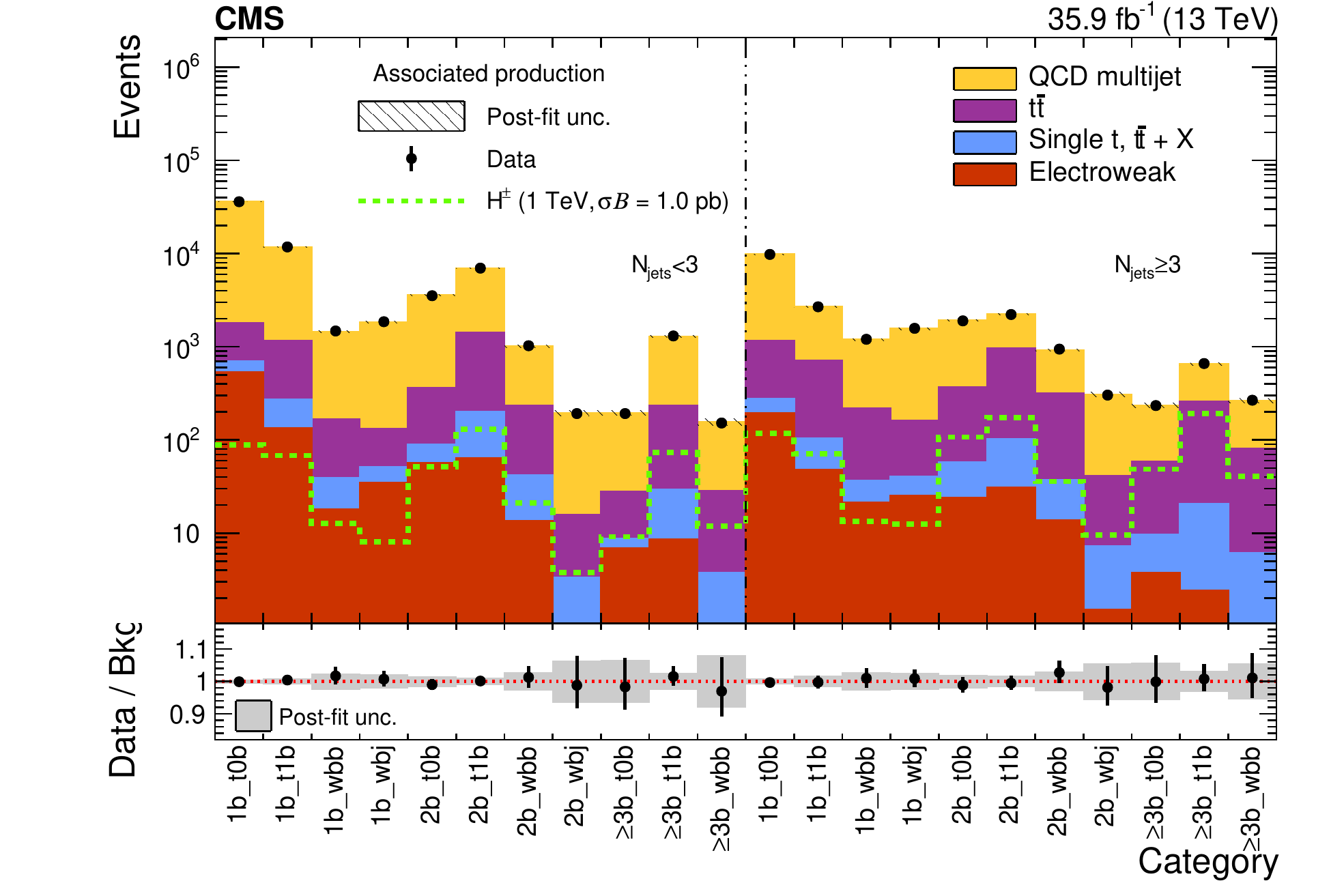}
\caption{\label{fig:YieldBoosted}
  Expected event yields for the \boostedAnalysis in the mass window as defined
  in Fig.~\ref{fig:MassCandidate_Signal} for an \Hpm with mass 1\TeV in each of the signal categories
  used in the associated production model. The 11 categories on
  the left have low jet multiplicity ($N_{\text{jets}} < 3$), while categories on
  the right have high jet multiplicity ($N_{\text{jets}} \ge 3$). The yields observed
  in data (black markers) are overlaid. The dashed lines represent the
  yields for an \Hpm with a mass of 1\TeV and $\sigma\mathcal{B}=1\pb$ for
  associated production. The background distributions result
  from the global fit described in the text for the background-only
  hypothesis. Similar categories are fitted for the $s$-channel production.
}

\end{figure}

\begin{figure}[htb]
\centering
\includegraphics[width=0.49\textwidth]{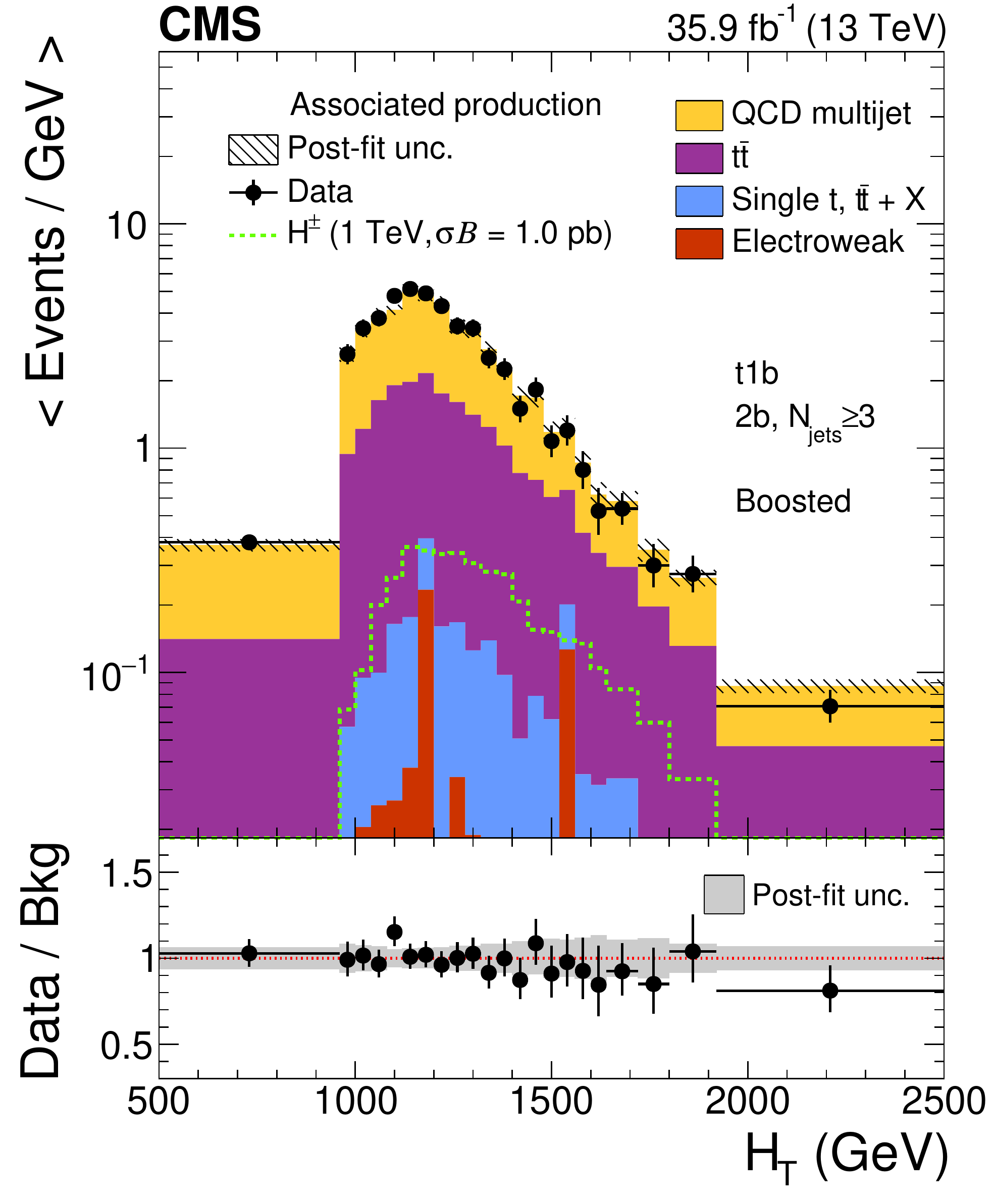}
\includegraphics[width=0.49\textwidth]{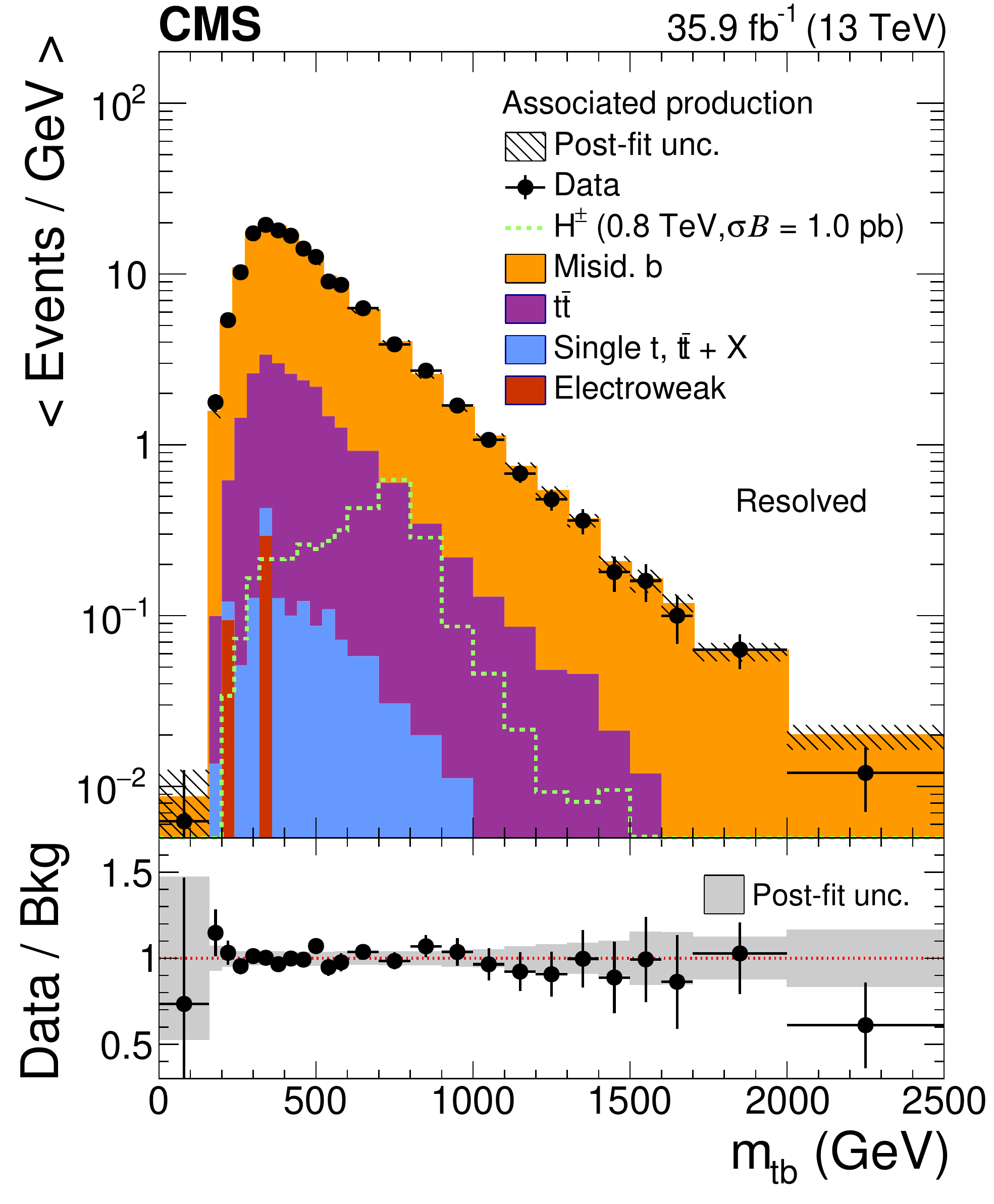}

\caption{\label{fig:fitvars}
  Variables used in the limit extraction. The \HT distribution is shown for the
  \boostedAnalysis and for the category t1b, 2b, $N_{\text{jets}} \geq 3$, in the mass window (\cmsLeft),
  for the associated production channels,
  with the expected signal for
  $\mHpm=1\TeV$. The invariant mass of the \PHpm candidates is shown
   for the
  \resolvedAnalysis (\cmsRight), with the expected signal for
  $\mHpm=0.8\TeV$. The background
  distributions result from the background-only fit discussed in the text.
  The distributions are binned according to
  the statistical precision of the samples,  leading to wider bins in
  the tail of the distributions.
  }

\end{figure}

The expected \SM event yields from a background-only fit to the data are
shown in \reffig{YieldBoosted} and \reftab{yields-resolved} for the boosted
and resolved analyses, respectively. For the \boostedAnalysis, the background
predictions are broken down into various categories of signal- and
background-enriched regions and in total 98 distributions are fitted.
The shape of the \HT distribution in the \boostedAnalysis and the invariant
mass of the \PHpm in the \resolvedAnalysis are
used to assess the agreement with the background-only hypothesis or the
presence of the signal in a global binned maximum likelihood fit incorporating
all the systematic uncertainties described in section~\ref{sec:systematics}
as nuisance parameters.
The fitted distributions for the background-only hypothesis are shown in
Fig.~\ref{fig:fitvars} (\cmsLeft) for
one category of the boosted analysis (t1b, 2b, $N_{\text{jets}} \ge 3$) and in
Fig.~\ref{fig:fitvars} (\cmsRight) for the resolved analysis.
The contribution of a hypothetical charged Higgs boson with a mass of 0.8
or 1\TeV and $\sigma\mathcal{B}=1\pb$, assuming the associated
production mechanism, is also displayed.

\begin{table}[!htb]
\centering
\topcaption{
  The numbers of expected and observed events for the \resolvedAnalysis
  after all selections. For background processes, the event yields and
  their corresponding uncertainties are prior to the background-only fit
  to the data. For the $\PSHpm$ mass hypotheses of 0.50, 0.65, and
  0.80\TeV, the signal yields are normalized to a
  $\sigma\mathcal{B}=1\pb$
  and the total systematic uncertainties prior to the fit are shown.
}
\label{tab:yields-resolved}
\renewcommand{\arraystretch}{1.2}
\begin{tabular}{ l c }
\multicolumn{1}{ l }{Process}  & Events $\pm$ (stat) $\oplus$ (syst)\\
\hline
Misidentified {\bjet}s    & $6152 \pm 292 $\\
\GenuineB & $1067 ~^{+ 185}_{- 187} $\\
Total expected from the SM              & $7220 \pm 336 $ \\
Observed &  7124 \\
$\PSHpm$ signal, $\mHpm = 0.5\TeV$ & $183 \pm 26 $\\
$\PSHpm$ signal, $\mHpm = 0.65\TeV$ & $218  ~^{+ 30}_{- 31} $\\
$\PSHpm$ signal, $\mHpm = 0.8\TeV$ & $234 \pm 33 $\\
\end{tabular}
\end{table}

\begin{figure}[htb]
\centering
  \includegraphics[width=0.49\textwidth]{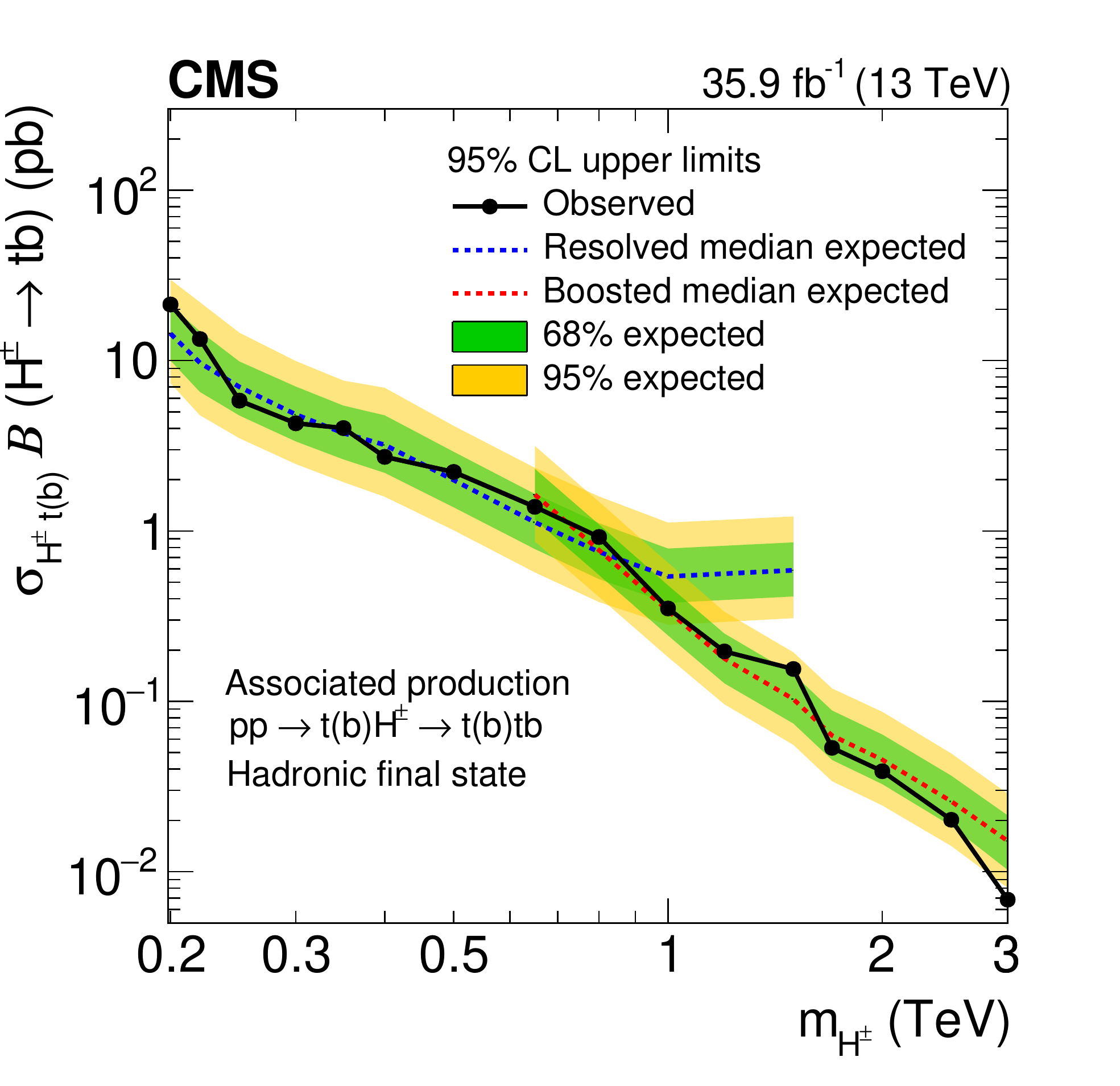}
  \includegraphics[width=0.49\textwidth]{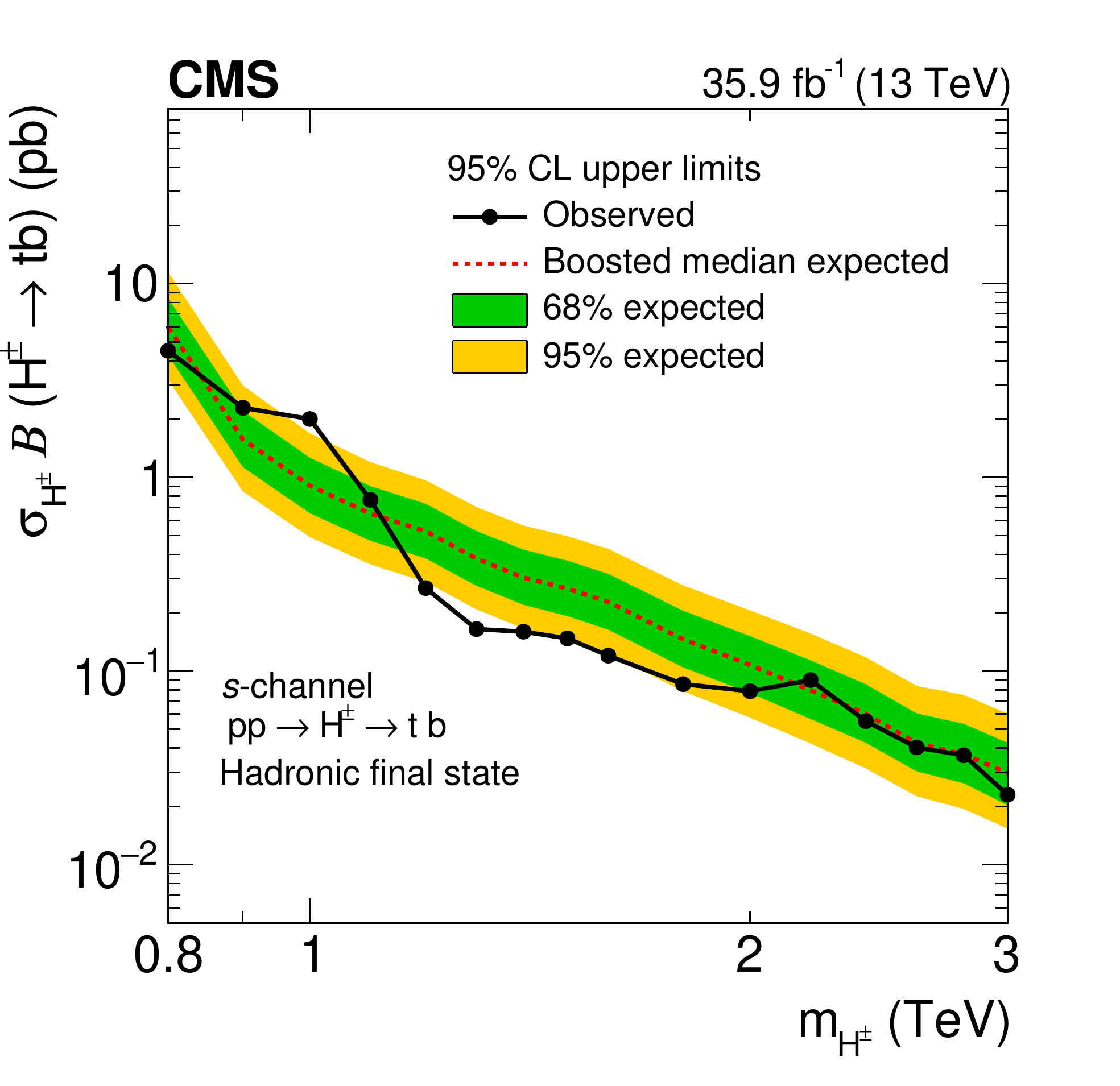}
  \caption{\label{fig:limits}
    Upper limits at 95\% \CL on the product of the \PHpm production cross section and branching
    fraction as a function of \mHpm for the associated (\cmsLeft) and
    $s$-channel (\cmsRight) processes. The observed upper limits are shown
    by the solid black markers. The median expected limit (dashed line),
    68\% (inner green band), and 95\% (outer yellow band) confidence
    interval for the expected limits are also shown.  For the association production model
    limits are calculated from the resolved (boosted) analysis for \mHpm
    points up to (greater than) 0.8\TeV.
  }

\end{figure}

The observed data agree with the predicted \SM background processes.
The results of the search are interpreted to set upper limits
on the product of the charged Higgs boson production cross section
and branching fraction into a top and bottom quark-antiquark pair.
The upper limits are calculated at 95\% \ConfLevel using the \CLs
criterion~\cite{Junk:1999kv,CLS2}. An asymptotic approximation is applied
for the test statistic~\cite{Cowan:2010js,COMB-NOTE},
${\text{ln}}\,\mathcal{L}_\mu/\mathcal{L}_{\text{max}}$,
where $\mathcal{L}_{\text{max}}$ is the maximum likelihood determined by
allowing all fitted parameters, including the signal
strength, $\mu$, to vary, and $\mathcal{L}_\mu$ is the maximum likelihood for a
fixed signal strength.
Results are shown for the associated production model in \reffig{limits}(\cmsLeft).
The reported limit at each mass value is determined by choosing the analysis
strategy (resolved or boosted) with the best expected sensitivity.
The data in the boosted analysis are also examined in the context of the
$s$-channel model and the resulting limits are shown in \reffig{limits}(\cmsRight).

Exclusion limits are placed on the production cross section of the \PHpm
associated with a top quark,
$\sPPtoHtb\BHtb=\sPPtoHptb\BHptb+\sPPtoHmtb\BHmtb$,
for masses from \massRangeAssCh\ in the range \xsecLimitAssCh.
The \boostedAnalysis has the best sensitivity for \mHpm larger than 0.8\TeV
while the \resolvedAnalysis limits are most stringent at lower masses.
The \boostedAnalysis sets upper limits from \xsecLimitSCh\ on the \PHpm
production cross section in the $s$-channel, $\sPPtoHpm\BHtb$,
for masses from \massRangeSCh, extending the regions excluded from prior 
results~\cite{Aad:2015typ}.

\begin{figure}[hbtp]
\centering
    \includegraphics[width=0.49\textwidth]{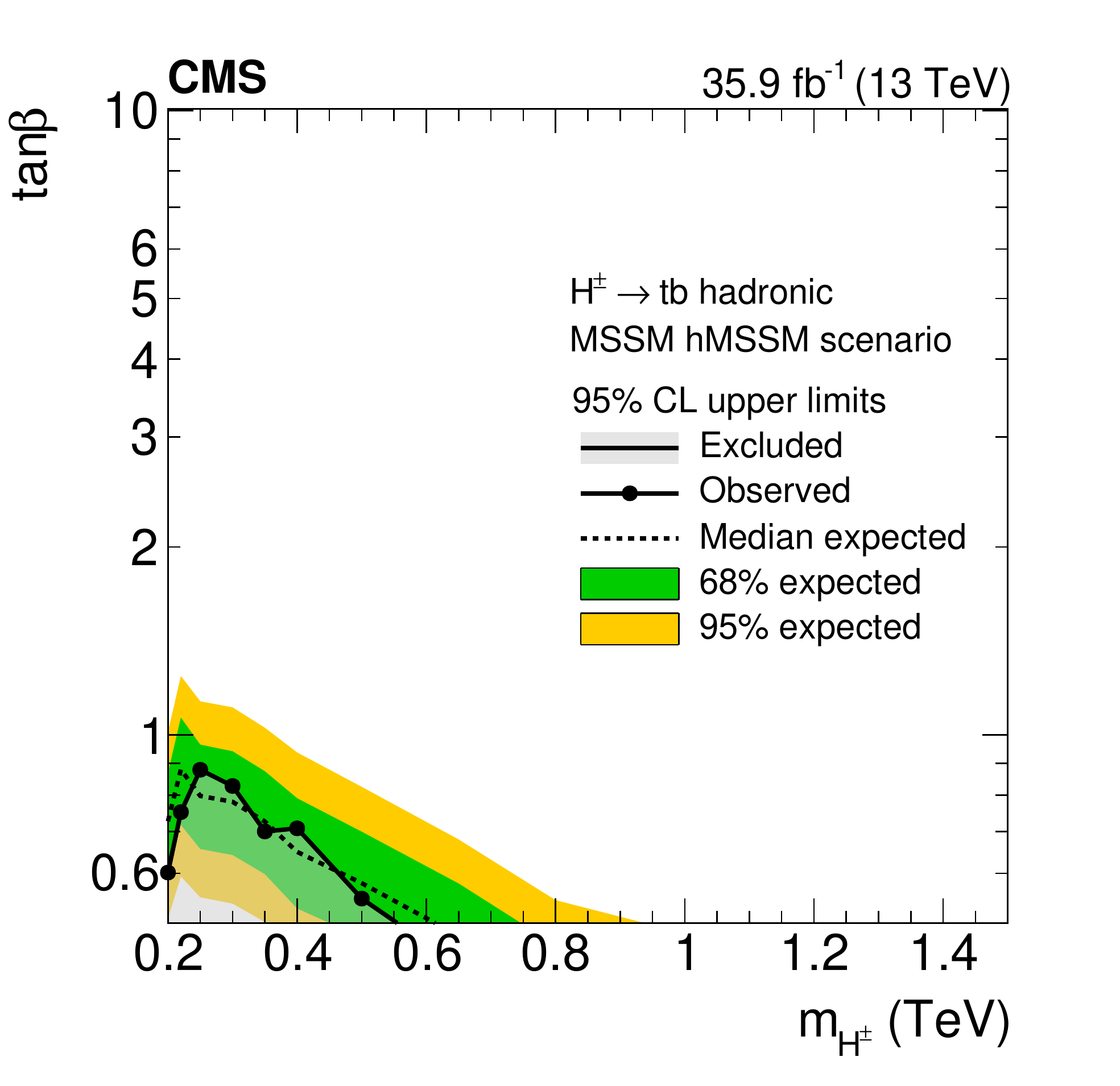}
    \includegraphics[width=0.49\textwidth]{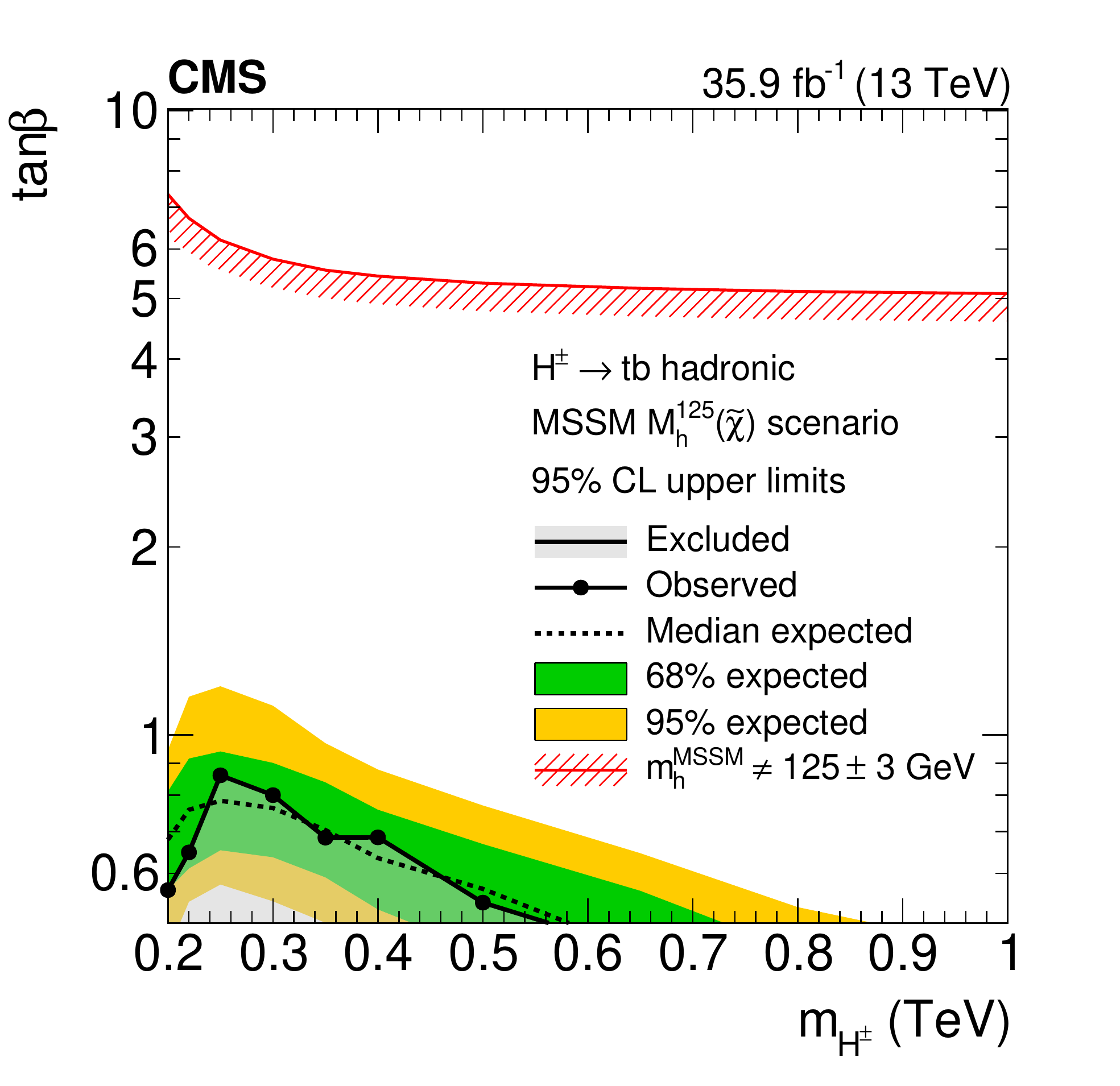}
    \caption{\label{fig:moddep}
        Excluded parameter space region in the hMSSM scenario (\cmsLeft) and
        \mhonetwentyfive (\cmsRight) using the association production model. The observed upper limits are shown by the
        solid black markers.  The median expected limit (dashed line),
        68\% (inner green band), and 95\% (outer yellow band) confidence interval
        for the expected limits are also shown.
        The region below the red line is excluded assuming that the observed neutral
        Higgs boson is the light CP-even \TwoHDM Higgs boson with a mass of ${125\pm 3\GeV}$,
        where the uncertainty is the theoretical uncertainty in the mass calculation.
    }
    
\end{figure}

Model-dependent upper limits are obtained by comparing the observed limit in the association production model
with theoretical predictions provided by the LHC-HXSWG~\cite{deFlorian:2016spz}.
The hMSSM benchmark scenario~\cite{Djouadi:2013vqa,Maiani:2013hud,Djouadi:2013uqa,Djouadi:2015jea}
assumes that the discovered Higgs boson is the light Higgs boson in the 2HDM and
that the SUSY particles have masses too large to be directly observed at
the LHC.
The \mhonetwentyfive{} scenario~\cite{Bahl:2018zmf} is
characterized as having significant mixing between higgsinos and gauginos, and a
compressed mass spectrum of charginos and neutralinos.
Its phenomenology differs from the Type II \TwoHDM due to
the presence of light charginos and neutralinos, such that heavy Higgs bosons are
allowed to decay to these superpartners.
Higgs masses and mixing in the \mhonetwentyfive{} scenario are computed
by \textsc{FeynHiggs}~\cite{Heinemeyer:1998yj,Heinemeyer:1998np,Degrassi:2002fi,Frank:2006yh,Hahn:2013ria,Bahl:2016brp,Bahl:2017aev}
in each point of the (\mA{}, \tanbeta) plane.
The branching fractions in the hMSSM scenario are calculated with
\textsc{hdecay}~\cite{Djouadi:1997yw,Djouadi:2006bz,Djouadi:2018xqq} alone,
while the \mhonetwentyfive{} scenario combines the most precise results
of \textsc{FeynHiggs}, \textsc{hdecay} and \textsc{PROPHECY4f}~\cite{Bredenstein:2006rh, Bredenstein:2006ha}.
Figure~\ref{fig:moddep} shows the excluded parameter space in these
\MSSM scenarios.
In the hMSSM scenario the maximum \tanbeta value excluded is 0.88 for
\mHpm values between 0.20 and 0.55\TeV. In the \mhonetwentyfive{} scenario
the maximum \tanbeta value excluded is 0.86 for \mHpm values between 0.20
and 0.57\TeV.

\section{Combination with the leptonic final states}
\label{sec:combination}

In Ref.~\cite{Sirunyan:2019arl} a search is presented for an \PHpm
with mass greater than that of the top quark and decaying into a top and
bottom quark-antiquark pair in the complementary leptonic final states.
Events are selected by the presence of a single isolated charged lepton
(\Pe or \Pgm) or an opposite-sign dilepton pair ($\Pe\Pe$, $\Pgm\Pgm$, $\Pe\Pgm$).
These are categorized according to the jet multiplicity and number of
\btagged jets and multivariate techniques are used to enhance the signal
and background discrimination in each category.  The search is based on
the same $\Pp\Pp$ collision data collected by the CMS experiment at a center-of-mass
energy of 13\TeV, corresponding to an integrated luminosity of 35.9\fbinv.

\begin{figure}[hbtp]
\centering
    \includegraphics[width=0.49\textwidth]{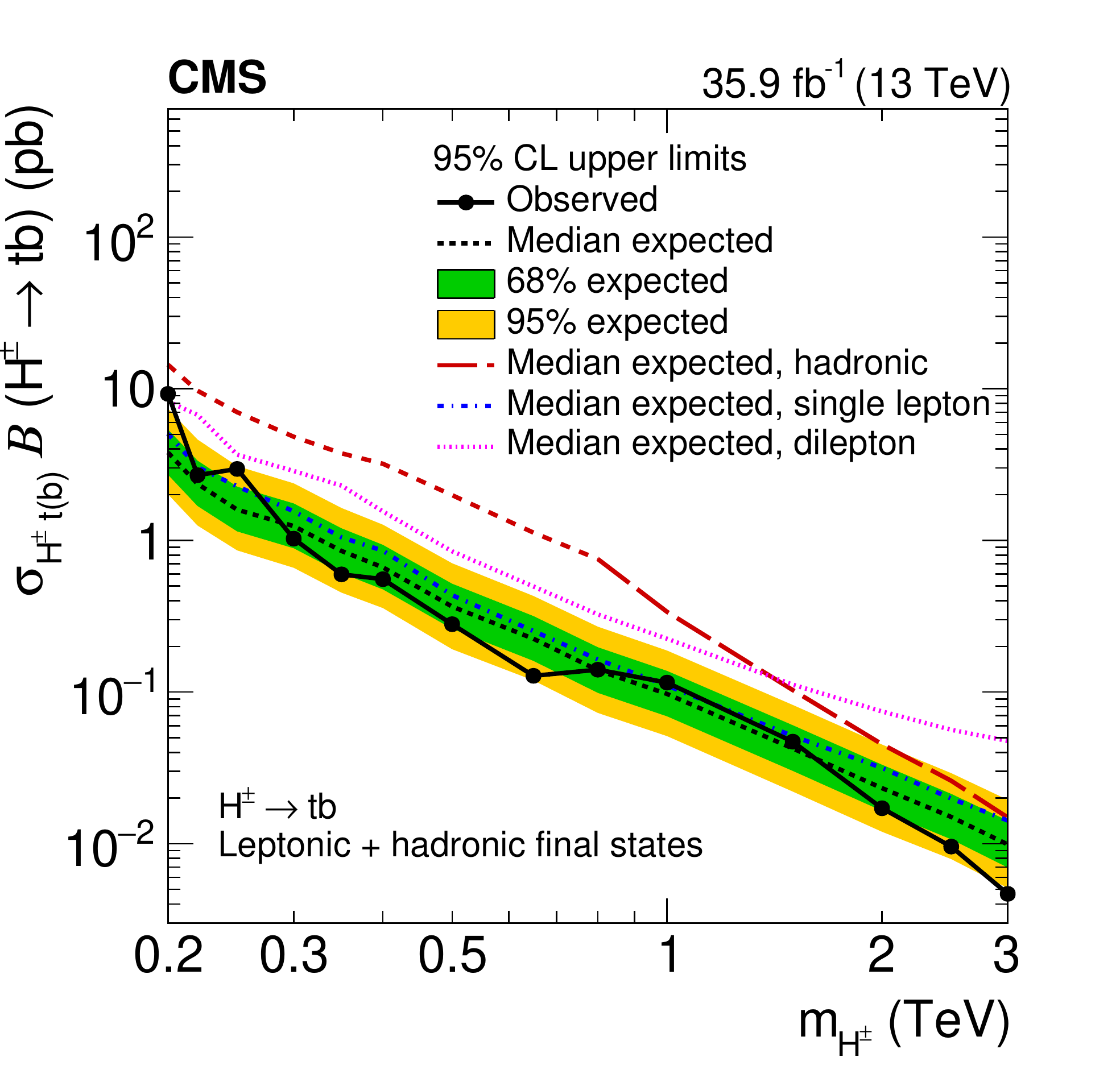}
    \includegraphics[width=0.49\textwidth]{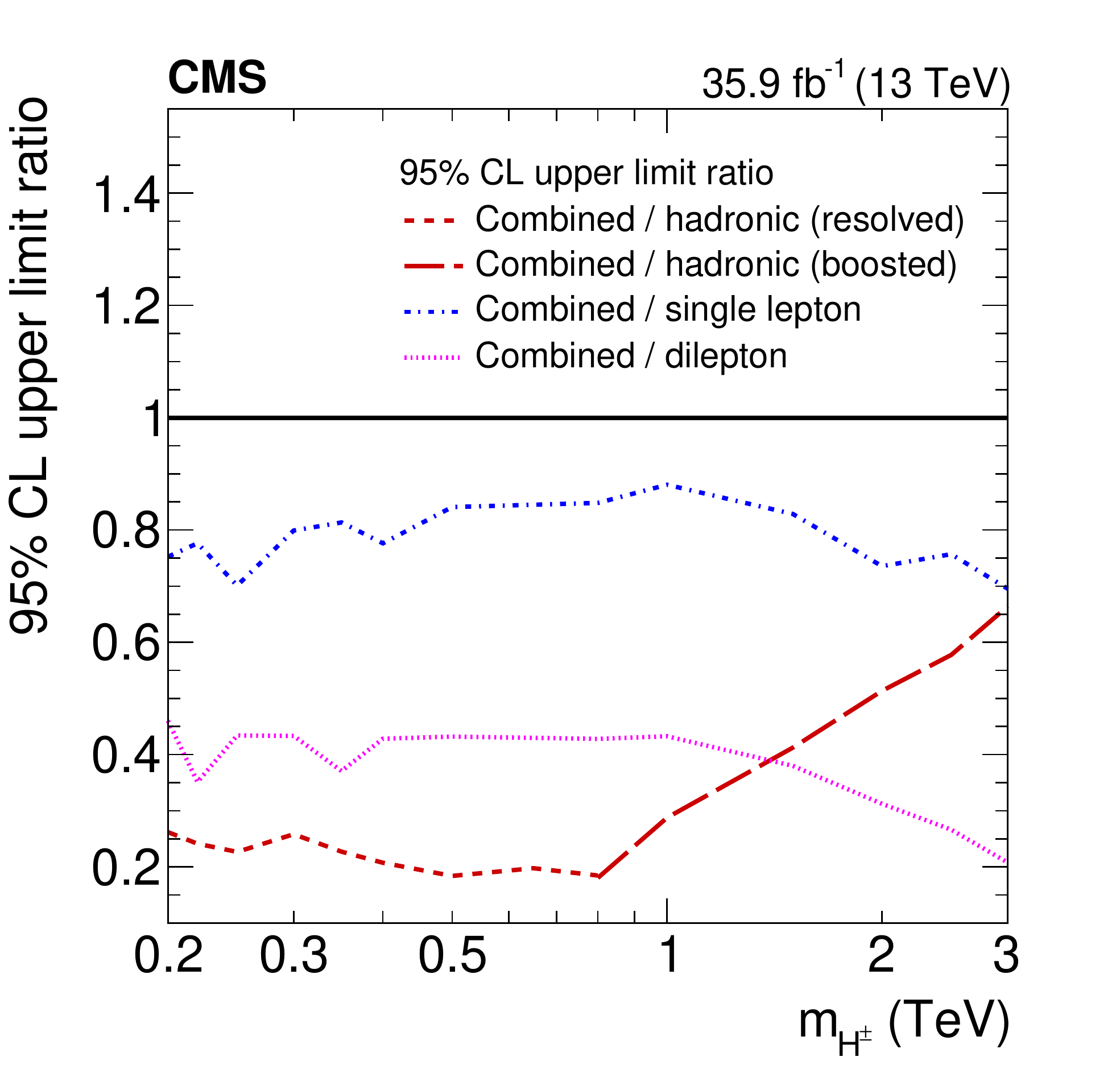}
    \caption{\label{fig:combination}
    Upper limits at 95\% \CL on the product of the \PHpm production cross section and branching
    fraction as a function of \mHpm for the process $\sPPtoHtb\BHtb$.
    The median expected limit (dashed line), 68\% (inner green band),
    and 95\% (outer yellow band) confidence interval expected limits are
    also shown (\cmsLeft). The relative expected contributions of each
    channel to the overall combination are shown (\cmsRight).
    The black solid line corresponds to the combined expected limits
    while the dashed, dotted and dash-dotted lines represent the contributing
    channels.
    }

\end{figure}

These results are combined with those from the all-jet channel
analyses to calculate the 95\% \CL combined upper limits on the
product of the cross section and the branching fraction as a function of
the \mHpm for the process $\sPPtoHtb\BHtb$. The limits are shown in
\reffig{combination} and Table~\ref{tab:Limit_Combined}.
The common experimental and theoretical nuisance parameters
between final states sharing the same production mechanism
are correlated, while the uncertainties from different sources described
in \refsec{systematics} are assumed to be uncorrelated.
The single-lepton final state has the best sensitivity in the whole
\mHpm range from \massRangeAssCh, while the dilepton channel contributes
in the low \mHpm regime, \ie, $\leq 1.5\TeV$, and the all-jet
channel improves the overall sensitivity by 20--25\% at larger values of \mHpm.

\begin{table}[htbp]
\centering
\topcaption{The upper limit at 95\% \CL on $\sPPtoHtb\BHtb$ with the combined all-jet, single-lepton, and dilepton final states.
}
\begin{tabular}{l c c c c c c}

\mc{}  & \multicolumn{5}{c}{Expected limits (pb)} & \mc{Observed limits (pb)} \\
\mc{\mHpm ({\TeVns})} & \mc{$-2$ s.d.} & \mc{$-1$ s.d.} & \mc{median}  &\mc{+1 s.d.}  & \mc{+2 s.d.} & \mbox{}\\
\hline

0.20 & 2.02 & 2.71 & 3.80 & 5.39 & 7.38 & 9.25 \\
0.22 & 1.25 & 1.69 & 2.36 & 3.36 & 4.62 & 2.69 \\
0.25 & 0.86 & 1.15 & 1.59 & 2.26 & 3.09 & 2.96 \\
0.30 & 0.66 & 0.89 & 1.24 & 1.75 & 2.38 & 1.03 \\
0.35 & 0.45 & 0.61 & 0.85 & 1.20 & 1.64 & 0.60 \\
0.40 & 0.36 & 0.48 & 0.66 & 0.93 & 1.27 & 0.56 \\
0.50 & 0.19 & 0.26 & 0.37 & 0.52 & 0.71 & 0.28 \\
0.65 & 0.12 & 0.16 & 0.22 & 0.32 & 0.43 & 0.13 \\
0.80 & 0.073 & 0.099 & 0.14 & 0.20 & 0.27 & 0.14 \\
1.00 & 0.051 & 0.069 & 0.097 & 0.14 & 0.19 & 0.12 \\
1.50 & 0.022 & 0.030 & 0.043 & 0.060 & 0.082 & 0.047 \\
2.00 & 0.012 & 0.017 & 0.023 & 0.033 & 0.045 & 0.017 \\
2.50 & 0.008 & 0.011 & 0.015 & 0.021 & 0.029 & 0.010 \\
3.00 & 0.005 & 0.007 & 0.010 & 0.014 & 0.019 & 0.005 \\

\end{tabular}
\label{tab:Limit_Combined}
\end{table}

\section{Summary}
\label{sec:summary}

Results are presented from a search for charged Higgs bosons (\PHpm)
that decay to a top and a bottom quark in the all-jet
final state. The search considers two distinct event topologies.
The \PHpm is reconstructed from a \btagged jet in combination with
a top quark candidate, either resolved as two jets from $\PQq\PAQq'$ decays
of a \PW boson and an additional \btagged jet, or, for highly boosted decay
products, reconstructed as a single top-flavored jet or a \PW jet
paired with an additional \btagged jet.
The analysis uses data collected with the CMS detector in 2016 at a
center-of-mass energy of \sqrts{13}, corresponding to an integrated
luminosity of \intLumi.  No significant deviation is observed above
the expected standard model background.
Model-independent upper limits at 95\% confidence level are set on the
product of the \PHpm production cross section and its branching fraction into a top and bottom quark-antiquark pair.
For production in association with a top quark, limits of \xsecLimitAssCh\
are set for \PHpm masses in the range \massRangeAssCh.
Combining these results with those from a search in leptonic final states
of $\PW$ bosons sets improved limits of \xsecLimitAssComb.
Exclusion regions are also presented in the parameter space of the
minimal supersymmetric
standard model hMSSM and \mhonetwentyfive{} benchmark scenarios.
The complementary $s$-channel
production of an \PHpm is investigated in the mass range \massRangeSCh\
and the corresponding upper limits are set at \xsecLimitSCh .

\begin{acknowledgments}
We congratulate our colleagues in the CERN accelerator departments for the excellent performance of the LHC and thank the technical and administrative staffs at CERN and at other CMS institutes for their contributions to the success of the CMS effort. In addition, we gratefully acknowledge the computing centers and personnel of the Worldwide LHC Computing Grid for delivering so effectively the computing infrastructure essential to our analyses. Finally, we acknowledge the enduring support for the construction and operation of the LHC and the CMS detector provided by the following funding agencies: BMBWF and FWF (Austria); FNRS and FWO (Belgium); CNPq, CAPES, FAPERJ, FAPERGS, and FAPESP (Brazil); MES (Bulgaria); CERN; CAS, MoST, and NSFC (China); COLCIENCIAS (Colombia); MSES and CSF (Croatia); RPF (Cyprus); SENESCYT (Ecuador); MoER, ERC IUT, PUT and ERDF (Estonia); Academy of Finland, MEC, and HIP (Finland); CEA and CNRS/IN2P3 (France); BMBF, DFG, and HGF (Germany); GSRT (Greece); NKFIA (Hungary); DAE and DST (India); IPM (Iran); SFI (Ireland); INFN (Italy); MSIP and NRF (Republic of Korea); MES (Latvia); LAS (Lithuania); MOE and UM (Malaysia); BUAP, CINVESTAV, CONACYT, LNS, SEP, and UASLP-FAI (Mexico); MOS (Montenegro); MBIE (New Zealand); PAEC (Pakistan); MSHE and NSC (Poland); FCT (Portugal); JINR (Dubna); MON, RosAtom, RAS, RFBR, and NRC KI (Russia); MESTD (Serbia); SEIDI, CPAN, PCTI, and FEDER (Spain); MOSTR (Sri Lanka); Swiss Funding Agencies (Switzerland); MST (Taipei); ThEPCenter, IPST, STAR, and NSTDA (Thailand); TUBITAK and TAEK (Turkey); NASU (Ukraine); STFC (United Kingdom); DOE and NSF (USA).

\hyphenation{Rachada-pisek} Individuals have received support from the Marie-Curie program and the European Research Council and Horizon 2020 Grant, contract Nos.\ 675440, 752730, and 765710 (European Union); the Leventis Foundation; the A.P.\ Sloan Foundation; the Alexander von Humboldt Foundation; the Belgian Federal Science Policy Office; the Fonds pour la Formation \`a la Recherche dans l'Industrie et dans l'Agriculture (FRIA-Belgium); the Agentschap voor Innovatie door Wetenschap en Technologie (IWT-Belgium); the F.R.S.-FNRS and FWO (Belgium) under the ``Excellence of Science -- EOS" -- be.h project n.\ 30820817; the Beijing Municipal Science \& Technology Commission, No. Z191100007219010; the Ministry of Education, Youth and Sports (MEYS) of the Czech Republic; the Deutsche Forschungsgemeinschaft (DFG) under Germany's Excellence Strategy -- EXC 2121 ``Quantum Universe" -- 390833306; the Lend\"ulet (``Momentum") Program and the J\'anos Bolyai Research Scholarship of the Hungarian Academy of Sciences, the New National Excellence Program \'UNKP, the NKFIA research grants 123842, 123959, 124845, 124850, 125105, 128713, 128786, and 129058 (Hungary); the Council of Science and Industrial Research, India; the HOMING PLUS program of the Foundation for Polish Science, cofinanced from European Union, Regional Development Fund, the Mobility Plus program of the Ministry of Science and Higher Education, the National Science Center (Poland), contracts Harmonia 2014/14/M/ST2/00428, Opus 2014/13/B/ST2/02543, 2014/15/B/ST2/03998, and 2015/19/B/ST2/02861, Sonata-bis 2012/07/E/ST2/01406; the National Priorities Research Program by Qatar National Research Fund; the Ministry of Science and Education, grant no. 14.W03.31.0026 (Russia); the Programa Estatal de Fomento de la Investigaci{\'o}n Cient{\'i}fica y T{\'e}cnica de Excelencia Mar\'{\i}a de Maeztu, grant MDM-2015-0509 and the Programa Severo Ochoa del Principado de Asturias; the Thalis and Aristeia programs cofinanced by EU-ESF and the Greek NSRF; the Rachadapisek Sompot Fund for Postdoctoral Fellowship, Chulalongkorn University and the Chulalongkorn Academic into Its 2nd Century Project Advancement Project (Thailand); the Kavli Foundation; the Nvidia Corporation; the SuperMicro Corporation; the Welch Foundation, contract C-1845; and the Weston Havens Foundation (USA).
\end{acknowledgments}

\bibliography{auto_generated}

\cleardoublepage \appendix\section{The CMS Collaboration \label{app:collab}}\begin{sloppypar}\hyphenpenalty=5000\widowpenalty=500\clubpenalty=5000\input{HIG-18-015-authorlist.tex}\end{sloppypar}
\end{document}

%% file: HIG-18-015-authorlist.tex
\vskip\cmsinstskip
\textbf{Yerevan Physics Institute, Yerevan, Armenia}\\*[0pt]
A.M.~Sirunyan$^{\textrm{\dag}}$, A.~Tumasyan
\vskip\cmsinstskip
\textbf{Institut f\"{u}r Hochenergiephysik, Wien, Austria}\\*[0pt]
W.~Adam, F.~Ambrogi, T.~Bergauer, J.~Brandstetter, M.~Dragicevic, J.~Er\"{o}, A.~Escalante~Del~Valle, M.~Flechl, R.~Fr\"{u}hwirth\cmsAuthorMark{1}, M.~Jeitler\cmsAuthorMark{1}, N.~Krammer, I.~Kr\"{a}tschmer, D.~Liko, T.~Madlener, I.~Mikulec, N.~Rad, J.~Schieck\cmsAuthorMark{1}, R.~Sch\"{o}fbeck, M.~Spanring, D.~Spitzbart, W.~Waltenberger, C.-E.~Wulz\cmsAuthorMark{1}, M.~Zarucki
% ...
\vskip\cmsinstskip
\dag: Deceased\\
1:  Also at Vienna University of Technology, Vienna, Austria\\
% ...
